\documentclass[12pt]{article}
\usepackage{a4wide}
\pdfoutput=1

\usepackage[utf8]{inputenc}
\usepackage{draft}
\usepackage{putex}
\usepackage[all]{xy}
\usepackage{stackrel,amssymb}
\usepackage{mathtools}
\usepackage{graphicx}
\usepackage{hhline}
\usepackage{cellspace}
\usepackage{caption}
\usepackage{dsfont}
\usepackage{amsfonts}
\usepackage{amsmath}
\usepackage{amsthm}
\usepackage{bm}
\usepackage{array}
\usepackage{subcaption}
\usepackage{epstopdf}
\usepackage{enumerate}
\usepackage{cite}
\usepackage{tensor}
\usepackage{slashed}
\usepackage{rotating}
\usepackage{multirow}
\usepackage{longtable}
\usepackage{relsize}
\usepackage{ytableau}
\usepackage{tikz,tikz-3dplot}
\usepackage{tikz-cd}
\usetikzlibrary{shapes.misc,shadows,fit,decorations.pathmorphing,positioning,trees,decorations.markings,decorations.pathreplacing,calc,shapes,patterns,arrows}
\usetikzlibrary{positioning,arrows}
\usetikzlibrary{decorations.pathmorphing,calc}
\usetikzlibrary{decorations.markings}
\newif\ifmirrorsemicircle
\usepackage{float}
\usepackage{tikz-cd} 
\usepgfmodule{shapes}
\usetikzlibrary{backgrounds, arrows,calc,shapes,decorations.pathreplacing, decorations.markings, automata,positioning}

\usepackage{environ}
\usepackage{listings}
\usepackage{subfiles}
\usepackage{tabularx}
\usepackage{footnote}
\usepackage{etoolbox}
\usepackage{thmtools}
\usepackage[framemethod=Tikz]{mdframed}
\usepackage{comment}
\usepackage[normalem]{ulem}
\usepackage{placeins}
\usepackage{makecell}
\usepackage[most]{tcolorbox}

\usepackage{xparse}
\usepackage{hyperref}
\usepackage[capitalize]{cleveref}
\usepackage{pdflscape}
\usepackage{everypage}

\definecolor{greenZS}{RGB}{50,153,83}
\definecolor{redZS}{RGB}{255,92,0}
\definecolor{blueZS}{RGB}{44,96,241}
\definecolor{grayZS}{RGB}{235,235,235}
\definecolor{orangeZS}{RGB}{255,133,0}

\crefname{definition}{Def.}{Defs.}

\allowdisplaybreaks[1] 

\ytableausetup{smalltableaux,centertableaux}

\hypersetup{
pdftitle={QCD large $N_f$},    
pdfauthor={\textcopyright\ Guillermo Arias Tamargo, Sergio Benvenuti, Diego Rodríguez Gómez},     
pdfsubject={HEP},   
pdfcreator={pdfLaTex},   
pdfproducer={LaTex}, 
pdfkeywords={},
colorlinks=true,
linkcolor=blue,
urlcolor=blue,
filecolor=blue,
citecolor=blue,
linktocpage=true
}

 \tikzset{
 	partial ellipse/.style args={#1:#2:#3}{
 		insert path={+ (#1:#3) arc (#1:#2:#3)}
 	}
 }

 \tikzset{
 	wave amplitude/.initial=0.2cm,
 	wave count/.initial=8,
 	mirror semicircle/.is if=mirrorsemicircle,
 	mirror semicircle=false,
 	wavy semicircle/.style={
 		to path={
 			let \p1 = (\tikztostart),
 			\p2 = (\tikztotarget),
 			\n1 = {veclen(\x2-\x1,\y2-\y1)},
 			\n2 = {atan2(\x2-\x1,\y2-\y1))} in
 			plot [
 			smooth,
 			samples=(\pgfkeysvalueof{/tikz/wave count}+0.5)*8+1, 
 			domain=0:1,
 			shift={($(\p1)!0.5!(\p2)$)}
 			] ({ 
 				(\x*180-\n2 + 180 + \ifmirrorsemicircle 1 \else -1 \fi * 90%
 			}:{ 
 			(%
 			\n1/2+\pgfkeysvalueof{/tikz/wave amplitude} * %
 			sin(
 			\x * 360 * (\pgfkeysvalueof{/tikz/wave count} + 0.5%
 			)%
 			)%
 		})
 	} (\tikztotarget)
 }
}

\pgfdeclaredecoration{complete sines}{initial}
{
	\state{initial}[
	width=+0pt,
	next state=sine,
	persistent precomputation={\pgfmathsetmacro\matchinglength{
			\pgfdecoratedinputsegmentlength / int(\pgfdecoratedinputsegmentlength/\pgfdecorationsegmentlength)}
		\setlength{\pgfdecorationsegmentlength}{\matchinglength pt}
	}] {}
	\state{sine}[width=\pgfdecorationsegmentlength]{
		\pgfpathsine{\pgfpoint{0.25\pgfdecorationsegmentlength}{0.5\pgfdecorationsegmentamplitude}}
		\pgfpathcosine{\pgfpoint{0.25\pgfdecorationsegmentlength}{-0.5\pgfdecorationsegmentamplitude}}
		\pgfpathsine{\pgfpoint{0.25\pgfdecorationsegmentlength}{-0.5\pgfdecorationsegmentamplitude}}
		\pgfpathcosine{\pgfpoint{0.25\pgfdecorationsegmentlength}{0.5\pgfdecorationsegmentamplitude}}
	}
	\state{final}{}
}

\tikzset{ 
	fermion/.style={thick, draw=mygreen, postaction={decorate}, decoration={markings, mark=at position .5 with {\arrow[mygreen]{triangle 45}}}} ,
	scalar/.style={thick, draw=blue, postaction={decorate}, decoration={markings, mark=at position .43 with {\arrow[blue]{triangle 45}}}} ,
	scalarr/.style={thick, draw=blue, postaction={decorate}, decoration={markings, mark=at position .56 with {\arrowreversed[blue]{triangle 45}}}},
	photon/.style={decorate, draw=black,thick,
		decoration={coil,amplitude=3.5pt, segment length=4pt}},
	photon1/.style={decorate, draw=black,thick,
		decoration={complete sines,amplitude=3pt, segment length=5pt}},
	photon2/.style={decorate, draw=black,thick,
		decoration={complete sines,amplitude=3pt, segment length=4pt}}
}

 \definecolor{mygreen}{RGB}{50,205,50}
 
 \makeatletter
 \newcommand*\bigcdot{\mathpalette\bigcdot@{.5}}
 \newcommand*\bigcdot@[2]{\mathbin{\vcenter{\hbox{\scalebox{#2}{$\m@th#1\bullet$}}}}}
 \makeatother
 
 \tikzset{
 	partial ellipse/.style args={#1:#2:#3}{
 		insert path={+ (#1:#3) arc (#1:#2:#3)}
 	}
 }

 \tikzset{
 	wave amplitude/.initial=0.2cm,
 	wave count/.initial=8,
 	mirror semicircle/.is if=mirrorsemicircle,
 	mirror semicircle=false,
 	wavy semicircle/.style={
 		to path={
 			let \p1 = (\tikztostart),
 			\p2 = (\tikztotarget),
 			\n1 = {veclen(\x2-\x1,\y2-\y1)},
 			\n2 = {atan2(\x2-\x1,\y2-\y1))} in
 			plot [
 			smooth,
 			samples=(\pgfkeysvalueof{/tikz/wave count}+0.5)*8+1, 
 			domain=0:1,
 			shift={($(\p1)!0.5!(\p2)$)}
 			] ({ 
 				(\x*180-\n2 + 180 + \ifmirrorsemicircle 1 \else -1 \fi * 90%
 			}:{ 
 			(%
 			\n1/2+\pgfkeysvalueof{/tikz/wave amplitude} * %
 			sin(
 			\x * 360 * (\pgfkeysvalueof{/tikz/wave count} + 0.5%
 			)%
 			)%
 		})
 	} (\tikztotarget)
 }
}

\pgfdeclaredecoration{complete sines}{initial}
{
	\state{initial}[
	width=+0pt,
	next state=sine,
	persistent precomputation={\pgfmathsetmacro\matchinglength{
			\pgfdecoratedinputsegmentlength / int(\pgfdecoratedinputsegmentlength/\pgfdecorationsegmentlength)}
		\setlength{\pgfdecorationsegmentlength}{\matchinglength pt}
	}] {}
	\state{sine}[width=\pgfdecorationsegmentlength]{
		\pgfpathsine{\pgfpoint{0.25\pgfdecorationsegmentlength}{0.5\pgfdecorationsegmentamplitude}}
		\pgfpathcosine{\pgfpoint{0.25\pgfdecorationsegmentlength}{-0.5\pgfdecorationsegmentamplitude}}
		\pgfpathsine{\pgfpoint{0.25\pgfdecorationsegmentlength}{-0.5\pgfdecorationsegmentamplitude}}
		\pgfpathcosine{\pgfpoint{0.25\pgfdecorationsegmentlength}{0.5\pgfdecorationsegmentamplitude}}
	}
	\state{final}{}
}

\tikzset{ 
	fermion/.style={thick, draw=greenZS, postaction={decorate}, decoration={markings, mark=at position .5 with {\arrow[greenZS]{triangle 45}}}} ,
	scalar/.style={thick, draw=blue, postaction={decorate}, decoration={markings, mark=at position .43 with {\arrow[blue]{triangle 45}}}} ,
	scalarr/.style={thick, draw=blue, postaction={decorate}, decoration={markings, mark=at position .56 with {\arrowreversed[blue]{triangle 45}}}},
	photon/.style={decorate, draw=mygreen,thick,
		decoration={complete sines,amplitude=3.5pt, segment length=4pt}},
	gluon/.style={decorate, draw=black,thick,
		decoration={coil,amplitude=3.5pt, segment length=4.5pt}},
	photon1/.style={decorate, draw=black,thick,
		decoration={complete sines,amplitude=3pt, segment length=5pt}},
	photon2/.style={decorate, draw=black,thick,
		decoration={complete sines,amplitude=3pt, segment length=4pt}}
}

\def\makeatletter{\catcode`\@=11}
\makeatletter
\def\mathbox#1{\hbox{$\m@th#1$}}%
\def\math@ccstyles#1#2#3#4#5#6#7{{\leavevmode
      \setbox0\mathbox{#6#7}%
      \setbox2\mathbox{#4#5}%
      \dimen@ #3%
      \baselineskip\z@\lineskiplimit#1\lineskip\z@
      \vbox{\ialign{##\crcr
             \hfil \kern #2\box2 \hfil\crcr
             \noalign{\kern\dimen@}%
             \hfil\box0\hfil\crcr}}}}
\def\mathaccstyles{\math@ccstyles\maxdimen}
\def\maththroughstyles{\math@ccstyles{-\maxdimen}}
\def\unity%
 {\maththroughstyles{.45\ht0}\z@\displaystyle {\mathchar"006C}\displaystyle 1}

\numberwithin{equation}{section}

\begin{document}

\begin{flushright}\footnotesize

\texttt{}
\vspace{0.01cm}
\end{flushright}

\mbox{}
\vspace{-0.5truecm}
\linespread{1.1}

\centerline{\LARGE \bf Infrared phases of 3d massless CS-QCD and large $N_f$}
\medskip

\vspace{1 truecm}

\centerline{
    { Guillermo Arias-Tamargo\textsuperscript{\textit{a,b}}} \footnote{guillermo.arias.tam@gmail.com}, Sergio Benvenuti\textsuperscript{\textit{c}} \footnote{benve79@gmail.com} and
    { Diego Rodr\'iguez-G\'omez\textsuperscript{\textit{a,b}}} \footnote{d.rodriguez.gomez@uniovi.es}}

\vspace{1cm}
\centerline{{\it  \textsuperscript{a}Department of Physics, Universidad de Oviedo}} \centerline{{\it C/ Federico Garc\'ia Lorca  18, 33007  Oviedo, Spain}}
\medskip
\centerline{{\it  \textsuperscript{b}Instituto Universitario de Ciencias y Tecnolog\'ias Espaciales de Asturias (ICTEA)}}\centerline{{\it C/~de la Independencia 13, 33004 Oviedo, Spain.}}
\medskip
\centerline{{\it  \textsuperscript{c}INFN, Sezione di Trieste, SISSA, Via Bonomea 265, 34136, Trieste, Italy}}
\medskip
\vspace{1cm}

\centerline{\small{\bf Abstract} }

\begin{center}
\begin{minipage}[h]{\textwidth}
We compute anomalous dimensions of quartic operators which are singlets under the $\mathrm{U}(N_f)$ global symmetry in Yang-Mills theories with Chern-Simons level $k$ in three dimensions coupled to $N_f$ Dirac fermions. In order to have analytic control, we consider  the regime $N_f\gg N_c\gg 1$, where the problem is reduced to the study of a flavor-adjoint and a flavor-singlet bilinears whose square give the quartic operators of interest. We provide evidence that these operators hit marginality, signaling instabilities which, for $\frac{2k}{N_f}\lsim 1$ suggest the spontaneous breaking of the global symmetry, and no symmetry breaking otherwise. For $k=N_f/2-1$ (the value corresponding to the domain walls of 4d QCD at $\theta=\pi$), the critical value $N_f^*$ is tantalizingly close to the lower end of the conformal window of QCD$_4$, suggesting a connection between conformal and global symmetry breaking in the 4d theory and in its domain walls. We also study, at $k=0$, other quartic operators containing a singlet when branched under $\mathrm{U}\left(\frac{N_f}{2}\right)\times \mathrm{U}\left(\frac{N_f}{2}\right)$, finding that they hit marginality precisely at the same point as their flavor-neutral cousins. Using the same technology we study bosonic CS-QCD$_3$, finding no hint of symmetry breaking where our analysis is applicable.

\end{minipage}

\end{center}
\newpage

\tableofcontents

\section{Introduction}

QCD in 3 dimensions (QCD$_3$) is very interesting both from a pure theoretical standpoint and because certain condensed matter systems may exhibit phase transitions described by an emergent non-abelian gauge field coupled to a number of fermion species (see \textit{e.g.} \cite{PhysRevB.65.165113}). In this paper we will consider QCD$_3$ with gauge group $\mathrm{SU}(N_c)$ coupled to $N_f$ 3d Dirac fermions in the fundamental representation of the gauge group. The 0-form flavor symmetry is $\mathrm{SU}(N_c)\times \mathrm{U}(1)_{\rm baryonic}$. One important difference with respect to its 4d counterpart is that the theory admits, in addition to the standard Yang-Mills kinetic term for the gauge field, a Chern-Simons (CS) term labelled by an integer $k$. 

On top the Yang-Mills interactions, the theory admits as relevant operators mass deformations for the matter fields. Choosing a flavor-preserving equal mass $m$ for all matter fields, Komargodski and Seiberg \cite{Komargodski:2017keh} conjectured the form of the phase space as a function of $m$ for fixed $N_c,\,N_f,\,k$. The upshot of the discussion is that, for $N_f>2k$, the low  energy dynamics is controlled by a TFQT which is the pure CS theory $\mathrm{SU}(N)_{k+\frac{N_f}{2}}$ for $m>m_0^*$ and $\mathrm{SU}(N)_{k-\frac{N_f}{2}}$ for $m<m_0^*$; with a phase transition separating both cases which could be first or second order. By tuning the bare mass of the matter fields, \cite{Komargodski:2017keh} defined the location of the phase transition at  $m_0^*=0$. In turn, for $N_f\leq 2k$, the low  energy dynamics is controlled by the pure CS theory $\mathrm{SU}(N)_{k+\frac{N_f}{2}}$ for $m>m^*$ and $\mathrm{SU}(N)_{k-\frac{N_f}{2}}$ for $m<-m^*$, with a quantum phase described by a sigma model with target space

\begin{equation}\label{KSsb}
\mathcal{M}=\frac{\mathrm{U}(N_f)}{\mathrm{U}\left(\frac{N_f}{2}+k\right)\times \mathrm{U}\left(\frac{N_f}{2}-k\right)}\,,
\end{equation}
opening up in the region $(-m^*,\,m^*)$. This picture was further sharpened and studied in \cite{Gaiotto:2017tne,Armoni:2019lgb,Argurio:2019tvw,Argurio:2020her}. 

In this paper we will examine this picture focusing on the point $m=0$ and studying CS-QCD$_3$ as a function of the parameters $(N_c,\,N_f,\,k)$. In principle, the ``experiment" one would like to do is to fix $N_c$ and vary $N_f$ and $k$. However, in order to have a handle into the problem, we consider the large $N_f$ regime, where, on general grounds, one expects the system to be described by a conformal field theory (CFT) which admits a systematic perturbative expansion in powers of $\frac{1}{N_f}$. Moreover, considering $N_c$ to be large as well (albeit with $\frac{N_c}{N_f}\ll 1$ in order not to enter the Veneziano regime) leads to simplifications. This allows to analytically search for instabilities of this CFT  in the form of relevant operators preserving the full global symmetry hitting marginality as $\frac{N_c}{N_f}$ and $\frac{k}{N_f}$ are varied.\footnote{These operators are often dubbed \textit{dangerously irrelevant}.} Barring fine-tunning, as these operators hit marginality and given that they preserve the same symmetry as the original theory, they should then be included in the action; generically triggering a flow towards a new IR theory. Hence, we expect these instabilities to give us a hint of the behavior of QCD$_3$ with vanishing fermion masses. The natural candidates for dangerously irrelevant operators are the irrelevant operators closest to marginality. In the case at hand, these are quartic operators in the fermion fields preserving the full $\mathrm{U}(N_f)$ symmetry. Motivated by this, in this paper we will compute, in the fixed point of QCD$_3$ at large $N_f$, the dimension of quartic fermionic operators which are singlets under the flavor symmetry as $\frac{N_c}{N_f}$ and $\frac{k}{N_f}$ are varied. Our main results are as follows:
\begin{itemize}
    \item We identify a line in the parameter space were a quartic fermion operator $(\psi^2_{\rm adj})^2$ hits marginality (where $\psi^2_{\rm adj}$ is the fermion bilinear transforming in the adjoint of $\mathrm{SU}(N_f)$). This line lies at values of $N_f/N_c \simeq 4 \sim 8$, so we expect our approximation $N_f \gg N_c$ to be qualitatively trustable.\footnote{In $3d$, it turns out the the large $N_f$ approximation is often pretty good all the way down to quite low values of $N_f$. For instance, in \cite{Benvenuti:2019ujm} the anomalous dimensions of various operators in minimally supersymmetric QED with $N_f$ flavors where matched with a dual theory at $N_f=2$, and a $10\%$ quantitative agreement was found. In any case, had the result be that the transition sits at $N_f/N_c \sim 1$, then we could not  draw any qualitative conclusion. This is what happens in the case of bosonic QCD in section \ref{sec:bosonic}.}
    We can give support to the existence of this line using the results at $k=0$ for the anomalous dimension of $\psi^2_{\rm adj}$ at second order in $1/N_f$ \cite{Gracey:2018fwq}. They signal that the point at which the quartic operator hits marginality happens at larger values of $N_f$ when taking into account more orders in the $1/N_f$ expansion.
    This is a strong indication that the quartic operators do hit marginality in a region of parameter space where the large $N_f$ approximation is qualitatively trustable.
    \item  We identify a line in the parameter space were a quartic fermion operator $(\psi^2_{\rm sing})^2$ hits marginality (where $\psi^2_{\rm sing}$ is the fermion bilinear transforming in the singlet of $\mathrm{U}(N_f)$). This line partially overlaps with that where $(\psi^2_{\rm adj})^2$ hits marginality at $N_f/N_c\simeq 2\sim 3$ and $k/2N_f\simeq 1$, but not for $k > 2N_f$. While further away from the regime of our approximation, we find the qualitative results to still be trustable.  
    \item The scenario proposed is that in the region inside the line where $(\psi^2_{\rm adj})^2$ hits marginality both conformal symmetry and global symmetry are broken. This is achieved by a vacuum expectation value for a Hubbard-Stratonovich field in the adjoint of the $\mathrm{SU}(N_f)$ symmetry. This is the quantum phase of Komargodski and Seiberg \cite{Komargodski:2017keh}.
    One nice consistency check of the scenario above is that if we assume the symmetry breaking pattern \eqref{KSsb}, the  Hubbard-Stratonovich field, as a $N_f \times N_f$ matrix must have $N_f/2-k$ positive eigenvalues and $N_f/2+k$ negative eigenvalues, so the flavors get massive $N_f/2-k$ positive masses and $N_f/2+k$ negative masses, and the CS level $k$ is shifted to zero. This implies that at low energies we get a trivial TQFT.\footnote{In general one would  expect a non trivial TQFT times the NLSM, as usually happens for instance in the massive phases of $3d$ $\mathcal{N}=1$ gauge theories.} On the other hand, in the region inside the line where $(\psi^2_{\rm sing})^2$ hits marginality, conformal symmetry is broken but the global symmetry remains intact.
    \item When $k= 1-N_f/2$, our $3$d QCD's are supposed to describe the domain wall of QCD in $4$d at $\theta=\pi$ \cite{Gaiotto:2017tne}. We find loose semi-quantitative evidence that $(N_f^*)_{4d}$ is equal to $(N_f)^*$ in $3$d. That is the lower bound of the conformal window of QCD$_4$ coincides with the lower bound of the conformal window of the $3$d theory living on its domain wall at $\theta=\pi$.
\end{itemize}

Let us stress that the problem of understanding the IR dynamics of 3d gauge theories has a long history. For the abelian case, dubbed QED$_3$, it has long been debated whether the flavor $\mathrm{U}(N_f)$ symmetry is spontaneously broken below a certain critical $N_f$ (see \cite{Pisarski:1984dj,Appelquist:1986fd, Appelquist:1988sr}, and also \cite{Kaveh:2004qa,Braun:2014wja,DiPietro:2015taa,Herbut:2016ide,Benvenuti:2018cwd,Benvenuti:2019ujm} for more recent references). One popular mechanism for this would-be phase transition is fixed point annihilation, where the de-stabilizing quartic operators are interpreted as signs of a near-by fixed point which, as the external parameters $(N_f,\,k)$ are tuned, collides and annihilates with the original theory resulting in loss of conformal invariance \cite{Kaplan:2010zz}. This possibility has been studied from the point of view of dynamical systems in \cite{Gukov:2016tnp} (see also \cite{Kuipers:2018lux}). The case of QCD$_3$ has also been long studied in the past (see \textit{e.g.} \cite{Appelquist:1989tc} for an early reference). More recently, it was considered in \cite{Gies:2005as,PhysRevB.78.054432, Goldman:2016wwk}, where in fact similar analysis to ours are carried. In recent times, another popular approach to the problem is to exploit consistency considerations across IR dualities to try to discern the IR phases. This is the approach of \cite{Komargodski:2017keh}, and it also has been fruitfully extended to QCD-like theories with other matter content \cite{Argurio:2019tvw,Gomis:2017ixy,Choi:2018tuh,Choi:2019eyl,Tanizaki:2018wtg,Armoni:2017jkl} (see also \cite{Lohitsiri:2022jyz,Bashmakov:2021rci,Delmastro:2021otj,Kan:2019rsz, Aitken:2019mtq,Akhond:2019ued,Amariti:2018wht,Dey:2018ykx,Aharony:2018pjn,Choi:2018ohn,Gaiotto:2018yjh,Benini:2018umh,Bashmakov:2018wts,Cordova:2017kue}).

The organization of this paper is as follows. In section \ref{sec:fermionic} we consider CS-QCD in 3d with $N_f$ Dirac fermions (for that reason we sometimes refer to it as fermionic QCD$_3$). After discussing the particularities of the CFT at large $N_f$, we turn to the discussion of the four flavor-singlet quartic operators which are natural candidates for dangerously irrelevant operators. We further consider $N_c$ to be large (yet small compared to $N_f$ in order not to destroy the approximation). This brings extra simplifications, since, due to large $N_c$ factorization, it turns out that the problem boils down to studying bilinear fermionic operators. There are two of these corresponding to the possible flavor structures: a flavor singlet and a flavor adjoint (consequently, the flavor-singlet quartic operators of interest are the square of a flavor singlet and flavor adjoint). We explicitly compute their dimension and check that, depending on the ratio $\frac{2k}{N_f}$, either the dimension of the flavor singlet or flavor adjoint goes towards marginality. We suggest that the locus when the corresponding operator hits marginality is resolved into fixed point annihilation. This allows us to draw a phase diagram in line with the expectations from the proposed phase diagram for QCD$_3$. To shed further light, we focus on $k=0$ and consider non-singlet operators under the $\mathrm{U}(N_f)$ flavor symmetry which contain a singlet when branched under the expected $\mathrm{U}\left(\frac{N_f}{2}\right)\times \mathrm{U}\left(\frac{N_f}{2}\right)$  after symmetry breaking. Reassuringly, we find that also those hit marginality at the precise same value of $\frac{N_f}{N_c}$ as their flavor singlet cousins. In section \ref{sec:bosonic} we apply the same tools to bosonic QCD$_3$. Finally, we wrap up in section \ref{sec:conclusions} with some conclusions. We collect in the appendices some details of the (more involved) computation of the scaling dimensions operators under study.

\section{Fermionic CS-QCD in 3d}\label{sec:fermionic}

As described above, our goal is to study QCD$_3$ with vanishing quark masses as a function of the external parameters $N_f,\,N_c,\,k$. To tackle the problem, we will consider the large $N_f$ limit, where the theory is at a fixed point and admits a systematic $\frac{1}{N_f}$ expansion.\footnote{Another popular strategy is to consider the $\epsilon$ expansion from $d=4$. In particular, \cite{Goldman:2016wwk} follows this approach to perform an analogous search to ours for instabilities in QCD$_3$.} We begin by briefly reviewing the large $N_f$ limit of QCD in which we will work, following \cite{Chester:2016ref}, and setting up the Feynman rules for the computation. The lagrangian is
\begin{align}
    \mathcal{L} = \frac{1}{4 g_{YM}^2} \text{Tr}\left[ F_{\mu\nu} F^{\mu\nu} \right] - \overline{\psi}^j_a \gamma^\mu \left( \delta_b^a \partial_\mu + i (T^A)^a_b A^A_\mu \right) \psi^b_j \,.
\end{align}
Here, $a$ and $b$ are colour indices in the fundamental representation, $A$ in the adjoint; and $j$ is a flavour index.

As advertised, in the large $N_f$ limit the theory flows to a CFT which can be systematically studied in a $\frac{1}{N_f}$ expansion. To understand how this comes about, note that in the large $N_f$ limit one should sum all the fermion bubble contributions to the gluon propagator as in Figure \ref{fig:effective_gluon}. 

\begin{figure}[h!]
	\centering
	\begin{tikzpicture} [scale=0.5]  
	\draw[gluon, red] (0,0) to (2,0);
	\node at (2.5,0) {$=$};
	\draw[gluon,black] (3,0) to (5,0);
	\node at (5.5,0) {$+$};
	\draw[gluon,black] (6,0) to (7.1,0);
	\draw[thick, blue] (8.1,0) circle(1cm);
	\draw[gluon,black] (9.1,0) to (10.2,0);
	\node at (10.7,0) {$+$};
	\draw[gluon,black] (11.3,0) to (12.4,0);
	\draw[thick, blue] (13.4,0) circle(1cm);
	\draw[gluon,black] (14.4,0) to (15.5,0);
	\draw[thick,blue] (16.5,0) circle (1cm);
	\draw[gluon,black] (17.5,0) to (18.6,0);
	\node at (19,0) {$ \ \ \  \ \ \ + \  \cdots$};
	\end{tikzpicture}
	\caption{Effective gluon propagator (red line). The black line stands for~the~bare~gluon~propagator, and the blue line for the fermion propagator.} \label{fig:effective_gluon}
\end{figure}
The (finite) contribution of each fermion bubble is

\begin{equation}
    \Pi^{\mu\nu}=\frac{N_f}{16}|p|\,\big(\delta^{\mu\nu}-\frac{p^{\mu}\,p^{\nu}}{p^2}\big)\,.
\end{equation}
Hence, choosing the Landau gauge

\begin{equation}
    (D^{\rm eff}_{\mu\nu})^{AB}=\frac{g_{YM}^2}{p^2\,(\frac{N_f\,g_{YM}^2}{16 |p|}-1)}\,\delta^{AB}\,\big(\delta_{\mu\nu}-\frac{p_{\mu}\,p_{\nu}}{p^2}\big)\,.
\end{equation}
At small $|p|$, the propagator becomes

\begin{equation}
    (D^{\rm eff}_{\mu\nu})^{AB}\sim\frac{16\,\delta^{AB}}{N_f\,|p|}\,\big(\delta_{\mu\nu}-\frac{p_{\mu}\,p_{\nu}}{p^2}\big)\,.
\end{equation}
Thus, in the large $N_f$ limit, the IR gluon propagator is dominated by the (finite) fermion bubbles. This in particular means that the YM kinetic term is dropped in the IR CFT. 

The conclusion above, while correct, heavily depends on the choice of gauge. In order to evade such gauge dependence, we use the same strategy as in \cite{Chester:2016ref}. In particular, we use a non-local gauge-fixing term in the action instead of the standard one,
\begin{align}\label{eq:nonlocal_gf}
    S_{g.f.}=\frac{N_f}{32(\xi-1)}\int d^3x\int d^3y\, \frac{\partial_\mu A_A^\mu(x)\,\partial_\nu A_A^\nu(y)}{2\pi^2|x-y|^2}\,.
\end{align}
Here $\xi$ is a gauge-fixing parameter which must drop at the end of the computations. Performing the resummation of Figure \ref{fig:effective_gluon} with this gauge fixing term results in an expression for the effective gluon propagator where one can indeed take the limit $g_{YM}^2\to\infty$ for any $\xi$, and is left with
\begin{align}\label{eq:gluon_without_CS}
\begin{tikzpicture}
\draw[gluon,red] (0,0) to (1.5,0);
\node at (4.1,0) { \scalebox{1.15}{$= \frac{16\, \delta^{AB}}{N_f|p|}\left(\delta_{\mu\nu}-\xi\frac{p_\mu p_\nu}{p^2}\right)\, .$}  };
\end{tikzpicture}
\end{align}
Note that, for $\xi=1$ this is precisely the same result as in the gauge-fixed argument above.

\begin{figure}[t!]
	\centering
	\begin{tikzpicture}[scale=0.4]
	
	\draw[black] (-4,3) to (21,3);
	\draw[black] (-4,3) to (-4,-30+2+10+5.5);
	\draw[black] (21,3) to (21,-30+2+10+5.5);
	\draw[black] (-4,-30+2+10+5.5) to (21,-30+10+2+5.5);
	
	\draw[gluon,red] (-2,0) to (2,0);
	\node at (-2,0.8) {\scriptsize \textcolor{red}{$A$}};
	\node at (2,0.8) {\scriptsize \textcolor{red}{$B$}};
	\node at (-2,-0.8) {\scriptsize \textcolor{red}{$\mu$}};
	\node at (2,-0.8) {\scriptsize \textcolor{red}{$\nu$}};
	\node at (11,0) {$=\frac{16 \, \delta^{AB}}{N_f (1+\lambda^2)|p|}\left(\delta_{\mu\nu}-\xi \frac{p_\mu p_\nu}{p^2}-\lambda \frac{p^\alpha}{|p|}\epsilon_{\alpha\mu\nu}\right)$};

	\draw[scalar,blue] (-2,-4) to (2,-4);
	\node at (-2,-3.5) {\scriptsize\textcolor{blue}{$a$}};
	\node at (2,-3.5) {\scriptsize\textcolor{blue}{$b$}};
	\node at (-2,-4.5) {\scriptsize\textcolor{blue}{$i$}};
	\node at (2,-4.5) {\scriptsize\textcolor{blue}{$j$}};
	\node at (5.1,-4) {$=i\frac{\slashed{p}}{p^2}\delta^{ab}\delta_{ij}$};
	
	\draw[gluon,red] (-2,-10+1) to (0.5,-10+1);
	\draw[scalar,blue] (2,-8.5+1) to (0.5,-10+1);
	\draw[scalar,blue] (0.5,-10+1) to (2,-11.5+1);
	\node at (-2,-8.2) {\scriptsize\textcolor{red}{$A$}};
	\node at (-2,-9.8) {\scriptsize\textcolor{red}{$\mu$}};
	\node at (1.25,-7.25) {\scriptsize\textcolor{blue}{$a$}};
	\node at (2.45,-7.75) {\scriptsize\textcolor{blue}{$i$}};
	\node at (1.25,-7.25-3-0.2) {\scriptsize\textcolor{blue}{$b$}};
	\node at (2.35,-10) {\scriptsize\textcolor{blue}{$j$}};
	\node at (6,-9) {$=i \gamma^\mu \, \delta^j_i\, (T^A)_a^b $};

	\end{tikzpicture}
	\caption{Summary of Feynman rules of fermionic QCD, after resumming the fermion bubbles in the gluon propagator. $A$ and $B$ are labels in the adjoint of the gauge group, while $a$ and $b$ label the fundamental representation. The flavour indices are $i,j$; while $\mu,\nu$ are spacetime indices as usual.} 
	\label{fig:fQCDfeynmanrules}
\end{figure}

At this point, adding a Chern-Simons term to the action is straightforward,\footnote{This is completely equivalent to the standard form $S_{CS}=\frac{k}{4\pi}\int \text{Tr}\left[ A\wedge dA+\frac{2}{3}A\wedge A\wedge A\right]$}
\begin{align}
    S_{CS}=\frac{k}{4\pi}\int d^3x\, A^A_{\mu}\epsilon^{\mu\nu\rho}F_{\nu\rho}^A\,,
\end{align}
and it leads to the standard additional term to the gluon propagator. It is worth recalling the well-known fact that the CS term gives a gauge-invariant mass to the gluon of the order $m_{\rm gluon}\sim g_{YM}^2\,k$. We would like to follow the same steps as before. In order to find the effective gluon propagator after resummation of the gluon bubbles, due to the unconventional coefficients in front of \eqref{eq:gluon_without_CS}, it is convenient to define
\begin{align}
    \lambda = \frac{8 k}{\pi N_f}\,,
\end{align}
which leads to \cite{Lee:2018udi}
\begin{align}
\begin{tikzpicture}
\draw[gluon,red] (0,0) to (1.5,0);
\node at (0.1,0.35) {\scriptsize \textcolor{red}{$A$}};
\node at (1.4,0.35) {\scriptsize \textcolor{red}{$B$}};
\node at (0.1,-0.35) {\scriptsize \textcolor{red}{$\mu$}};
\node at (1.4,-0.35) {\scriptsize \textcolor{red}{$\nu$}};
\node at (5.6,0) { \scalebox{1.15}{$= \frac{16\, \delta^{AB}}{N_f(1+\lambda^2)|p|}\left(\delta_{\mu\nu}-\xi\frac{p_\mu p_\nu}{p^2} - \lambda\frac{p^\alpha}{|p|}\epsilon_{\alpha\mu\nu}\right)\, .$}  };
\end{tikzpicture}
\end{align}
For ease of reference, we have summarized all the resulting Feynman rules in Figure \ref{fig:fQCDfeynmanrules}. Sometimes we will also use the fermion propagator in position space, which equals
\begin{align}
    G(x,0) = \delta_i^j\delta_a^b\,\frac{x_\mu \gamma^\mu}{4 \pi |x|^3}\,.
\end{align}

\subsection{Flavor-singlet quartic operators}\label{sec:nearby_fixed_points}

The natural candidates for dangerously irrelevant operators destabilizing the large $N_f$ fixed point of QCD$_3$ are the irrelevant operators closest to marginality, which in 3d are those of dimension 4. There are four such independent operators, which we can take to be the following quartic fermion operators:\footnote{Other operators with the same classical scaling dimension, involving $F^2$, are redundant, in the sense that they can be written in terms of the four quartic ferminic operators using the equations of motion \cite{Chester:2016ref,Goldman:2016wwk}.}
\begin{align}
  \mathcal{O}_1 = \left( \overline{\psi}^i_a \, \psi_i^a\right) \, \left(\overline{\psi}^j_b \, \psi_j^b \right)\,,\\
  \mathcal{O}_2 = \left( \overline{\psi}^i_a \, \psi_j^a \right)\, \left(\overline{\psi}^j_b \, \psi_i^b \right)\,,\\
  \mathcal{O}_3 = \left( \overline{\psi}^i_a \, \psi_i^b\right) \, \left(\overline{\psi}^j_b \, \psi_j^a\right)\,, \\
  \mathcal{O}_4 = \left( \overline{\psi}^i_a \, \psi_j^b\right) \, \left(\overline{\psi}^j_b \, \psi_i^a \right)\,.
\end{align}

Our task is to compute the anomalous dimension of these operators in the large $N_f$ limit. This is a complicated task which is greatly simplified if we further assume $N_c$ to large (yet with $N_c\ll N_f$ in order not to enter the Veneziano regime). This has important consequences, as we now describe. On general grounds, we can expect that there will be some complicated mixing between all of the quartic operators above. Note, however, that two of these operators are double trace in the color indices, while the other two are single trace. All diagrams that contribute to the correlation functions  between single trace operators, as well as the mixing between the single trace and double trace, will contain less powers of $N_c$ when compared to the correlator of two double trace operators. Hence, in the large $N_c$ limit the computation simplifies greatly as we can only consider $\mathcal{O}_1$ and $\mathcal{O}_2$. Furthermore, in this regime we also have the factorization property of correlation functions, which further simplifies the computation of the anomalous dimension of the double trace operators. More concretely, let us denote
\begin{align}
    &\psi^2_{\text{adj}}(x)=\overline{\psi}^i_{a}\psi^{a}_j-\frac{\delta_i^j}{N_f}\sum_k\overline{\psi}_{a}^k\psi^{a}_k\, ,\label{eq:psi2adj}\\
    &\psi^2_{\text{sing}}(x)=\frac{1}{\sqrt{N_f}}\sum_k\overline{\psi}^k_a\psi^{a}_k\, .\label{eq:psi2sing}
\end{align}
These two bilinear operators transform in the adjoint and the singlet of the flavour symmetry respectively, and their squares are the quartic operators $\mathcal{O}_1$ and $\mathcal{O}_2$ above. Then, factorization implies that
\begin{align}
    &\Delta[\mathcal{O}_1] = 2 \Delta[\psi^2_{\text{sing}}]\,,\\
    &\Delta[\mathcal{O}_2] = 2 \Delta[\psi^2_{\text{adj}}]\,.
\end{align}

In conclusion, working in the $N_f\gg N_c\gg 1$ regime, we reduce the problem of computing the anomalous dimension of four quartic operators to two bilinear ones; and moreover, as these two live in different representations of the global symmetry, we have the guarantee that they cannot mix. 

\begin{figure}[b!]
	\centering
	\begin{tikzpicture}[scale=0.4]
	
	\draw[black] (-4,3) to (20,3);
	\draw[black] (-4,3) to (-4,-30+2+10);
	\draw[black] (20,3) to (20,-30+2+10);
	\draw[black] (-4,-30+2+10) to (20,-30+10+2);
	
	\draw[thick, blue] (0,0) circle (2cm);
	\draw[thick,fill] (-2,0) circle (4pt);
	\draw[thick,fill] (2,0) circle (4pt);
	
	\node at (5.9,0) {$=-\frac{N_fN_c}{8\pi^2x^4}$};
	
	\draw[blue,thick] ([shift=(180:2cm)]0,-5)   arc (180:0:2cm);
	\draw[blue,thick] ([shift=(0:2cm)]0,-5)   arc (0:-180:2cm);
	\draw[gluon,red] (-1.66,-3.9) to (1.66,-3.9);
	\draw[thick,fill] (-2,-5) circle (4pt);
	\draw[thick,fill] (2,-5) circle (4pt);
	\node at (10.3,-5) {$ =\frac{(N_c^2-1)}{\pi^4(1+\lambda^2)x^4}\left(\frac{1}{3}-\xi\right)\log x^2\Lambda^2$};
	
	\draw[blue,thick] ([shift=(180:2cm)]0,-10)   arc (180:0:2cm);
	\draw[blue,thick] ([shift=(0:2cm)]0,-10)   arc (0:-180:2cm);
	\draw[gluon, red] (0,-10) ([shift=(90:2cm)]0,-10)  to   ([shift=(-90:2cm)]0,-10);
	\draw[thick,fill] (-2,-10) circle (4pt);
	\draw[thick,fill] (2,-10) circle (4pt);
	\node at (10.5,-10) {$=-\frac{(N_c^2-1)}{\pi^4(1+\lambda^2)x^4}(3-\xi)\log x^2\Lambda^2 $}; 
	
	\draw[thick,blue]  ([shift=(0:2cm)]0,-15)  arc (0:180:2cm);
	\draw[thick,blue]  ([shift=(-180:2cm)]0,-15) arc (-180:0:2cm); 
	\draw[gluon, red] ([shift=(40-14:2cm)]0,-15)  to   ([shift=(140+14:2cm)]6,-15);
	\draw[gluon, red] ([shift=(-40+14:2cm)]0,-15)  to   ([shift=(-140-14:2cm)]6,-15);
	\draw[thick,blue ] ([shift=(180:2cm)]6,-15) arc (180:0:2cm);
	\draw[thick,blue] ([shift=(0:2cm)]6,-15) arc (0:-180:2cm);
	\draw[thick,fill] (-2,-15) circle (4pt);
	\draw[thick,fill] (8,-15) circle (4pt);
	\node at (14,-15) {$=-\frac{8(N_c^2-1)(\lambda^2-1)}{\pi^4(1+\lambda^2)^2x^4}\log x^2\Lambda^2$};
	
	\end{tikzpicture}
	\caption{ (fQCD) Results for individual Feynman diagrams appearing in the 2-point correlation function of the fermion-bilinear operators.} \label{fQCDbilinears}
\end{figure}

In order to compute the desired anomalous dimensions, note that the two point function of each of these operators has the form
\begin{align}
    \langle \mathcal{O} \mathcal{O} \rangle = \frac{c_0}{|x|^{2\Delta}} \sim \frac{c_0}{|x|^{2\Delta_{\rm cl}}}-\frac{\gamma}{N_f}\frac{c_0}{|x|^{2\Delta_{\rm cl}}} \,\log(\Lambda^2|x|^2)+\cdots\,;
\end{align}
where we have split $\Delta=\Delta_{\rm cl}+\frac{\gamma}{N_f}$. Thus, we can read-off the anomalous dimension as 
\begin{align}
    \frac{\gamma}{N_f} = \frac{c_{\log}}{c_0}\,,
\end{align}
where $c_{\log}$ is the coefficient of $-\log\Lambda^2$. Hence, although the computation of the relevant diagrams for the 2-point functions can be highly involved, for our purposes it is enough to isolate the logarithmic divergences, as the anomalous dimension can be directly read off from their coefficient. When the dust settles, the results of the logarithmic divergence of each diagram are those in Figure \ref{fQCDbilinears} (the technical details of the computations can be seen in appendix \ref{app:computation_singlets}). Using these, the resulting scaling dimensions of the two operators under study are
\begin{align}\label{eq:fqcd_result_anomalous_dimension}
   & \Delta[\psi^2_{\text{adj}}]=2-\frac{64(N_c^2-1)}{3\pi^2N_c(1+\lambda^2)N_f}+O\left(\frac{1}{N_f^2}\right)\, ,\\
   & \Delta[\psi^2_{\text{sing}}]=2-\frac{128(N_c^2-1)(2\lambda^2-1)}{3\pi^2N_c(1+\lambda^2)^2 N_f}+O\left(\frac{1}{N_f^2}\right)\, .
\end{align}

\subsection{Fixed point merging and phase diagram}

\begin{figure}[h!]
    \centering
    \includegraphics[width=0.5\textwidth]{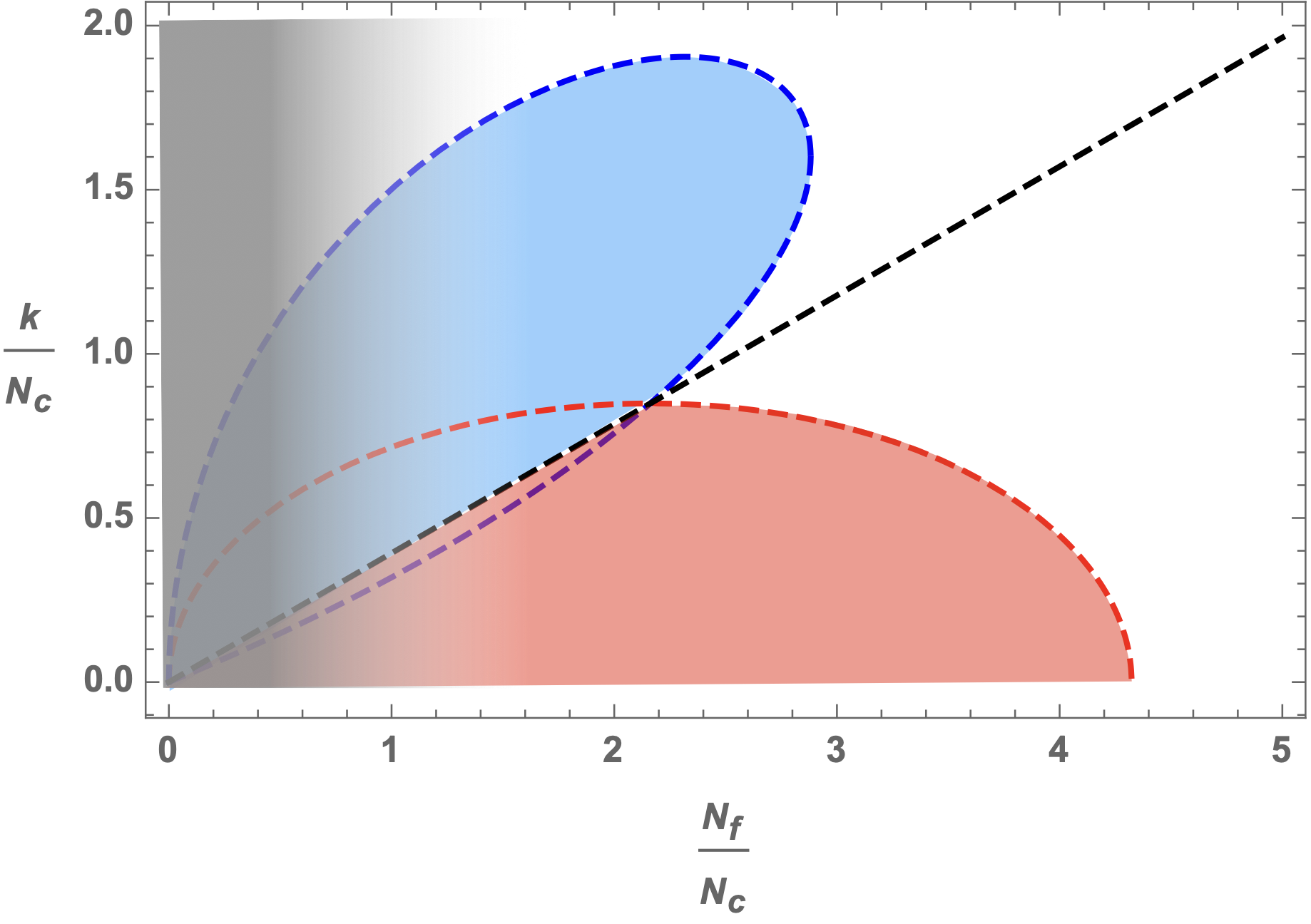}
    \caption{Locus in parameter space where the operators $(\psi^2_{\text{adj}})^2$ (red dashed curve) and $(\psi^2_{\text{sing}})^2$ (blue dashed curve) hit marginality. The line $\frac{N_f}{2k}=\frac{4}{\pi}$ is drawn in black. The shaded grey area represents the region of $N_f\sim N_c$ where our results cannot be extrapolated.}
    \label{fig:phase_diagram_fermionicQCD}
\end{figure}

Now that we have computed the scaling dimension of $\psi^2_{\text{adj}}$ and $\psi^2_{\text{sing}}$ as a function of $\frac{N_c}{N_f}$ and $\frac{k}{N_f}$, we can study the resulting phase diagram for QCD$_3$. From the results \eqref{eq:fqcd_result_anomalous_dimension}, and according to the argument in section \ref{sec:nearby_fixed_points}, the scaling dimension of the two fermionic quartic operators in the $N_f \gg N_c\gg 1$ regime are
\begin{align}
   &\Delta[\mathcal{O}_1]= \Delta[\left(\psi^2_{\text{adj}}\right)^2]=4-\frac{128}{3\pi^2(1+\lambda^2)}\,\frac{N_c}{N_f}+\cdots\, ,\\
   &\Delta[\mathcal{O}_2]= \Delta[\left(\psi^2_{\text{sing}}\right)^2]=4+\frac{128}{3\pi^2(1+\lambda^2)}\,\frac{N_c}{N_f}\,\frac{2(1-2\lambda^2)}{(1+\lambda^2)}+\cdots\, .
\end{align}
From this we see that for small values of $\frac{k}{N_f}$ corresponding to $\lambda^2<\frac{1}{2}$, the anomalous dimension of $\mathcal{O}_2$ is positive while that of $\mathcal{O}_1$ is negative. In particular, as $\frac{N_c}{N_f}$ is increased at fixed $\lambda$, $\Delta[\mathcal{O}_1]$ decreases towards marginality, thus signaling an \textit{a priori} instability. As $\frac{k}{N_f}$ is increased, the anomalous dimension of $\mathcal{O}_2$ becomes negative and eventually is such that $\Delta[\mathcal{O}_2]$ decreases faster than $\Delta[\mathcal{O}_1]$ towards marginality. This change of behavior happens at $\lambda=1$, which translates into $\frac{N_f}{2k}=\frac{4}{\pi}\sim 0.78$. The result of the analysis is compiled in Figure \ref{fig:phase_diagram_fermionicQCD} below (note that we choose to plot $\frac{k}{N_c}$ \textit{vs.} $\frac{N_f}{N_c}$).

The region in white in Figure  \ref{fig:phase_diagram_fermionicQCD} is described by the large $N_f$ CFT. As the parameters are varied towards the dashed curves, the corresponding operator decreases its dimension hitting marginality right at the curve, thus becoming a dangerously irrelevant operator and showing an instability. When that happens such operator must be included in the lagrangian and it will generically trigger a further RG flow towards a new IR theory.

\subsubsection{Instabilities as fixed point annihilation}

An alternative point of view on dangerously irrelevant operators comes from observing that an operator approaching marginality suggests, in conformal perturbation theory, the existence of a nearby fixed point --in this case commonly denoted by QCD$_3^*$. To make this explicit we may consider
\begin{equation}
\label{eq:SQCD*}
S_{{\rm QCD}_3^*}=S_{{\rm QCD}_3}+h\,\int \mathcal{O}\,,
\end{equation}
where $S_{{\rm QCD}_3}$ is the action for the CS-QCD$_3$ fixed point and $\mathcal{O}$ is the quartic operator in the matter fields in question (in our case, 
either $\mathcal{O}_1$ for $\frac{2k}{N_f}<0.78$ or $\mathcal{O}_2$ for $\frac{2k}{N_f}>0.78$). On general grounds, the beta function for $h$ is

\begin{equation}
\label{eq:betalambda}
\beta_{h} = \Big(d-\Delta\Big)\,h-b\,h^2+\cdots\,,
\end{equation}
where $b$ is some numerical coefficient coming from loop effecs. We should stress that the dimension of the operator $\mathcal{O}$ in the QCD fixed point, $\Delta$, depends on the values of the parameters $\frac{k}{N_f},\,\frac{N_c}{N_f}$. 

Besides the CS-QCD$_3$ fixed point at $h^{\star}=0$, the beta function has another fixed point at 

\begin{equation}
h^{\star}\sim \frac{\Delta-d}{b}\,.
\end{equation}
Thus, provided that $\Delta\sim d$ --so that the perturbative computation is valid--, this shows that there is another (unstable) fixed point --QCD$_3^*$-- which differs from QCD in the quartic operator $\mathcal{O}$. Moreover, these two fixed points approach each other as $\frac{k}{N_f},\,\frac{N_c}{N_f}$ approach their critical values for which $\Delta=d$.

To understand what happens in the region around the fixed point crossing it is necessary to take into account higher order corrections to \eqref{eq:betalambda}. In particular, due to gluon exchange graphs, one would expect an extra term in $\beta_{h}$ of the form $-c$, with $c$ a ``constant" (in that it does not depend on $h$) which depends on $\frac{N_c}{N_f}$ and $\frac{k}{N_f}$ and which we expect to be positive (this is so in QED$_3$ --see \textit{e.g.} \cite{Braun:2014wja} and the discussion in section 4 of \cite{Gukov:2016tnp}-- and in 4d QCD \cite{Kaplan:2010zz}). That is,
\begin{equation}
\label{eq:betalambdaUNFOLDED}
\beta_{h} = \Big(d-\Delta\Big)\,h-b\,h^2-c+\cdots\,.
\end{equation}
This last term in \eqref{eq:betalambdaUNFOLDED} has the effect of ``resolving'' the crossing and turning it into fixed point annihilation \cite{Gukov:2016tnp} as shown in Figure  \ref{fig:fixed_point_annhiliation} below. It should be noted that if $c$ had the opposite sign, rather than fixed point annihilation one would have that the fixed points cross without touching (see \textit{e.g.} \cite{Gukov:2016tnp}).

\begin{figure}[h!]
    \centering
    \includegraphics[width=0.8\textwidth]{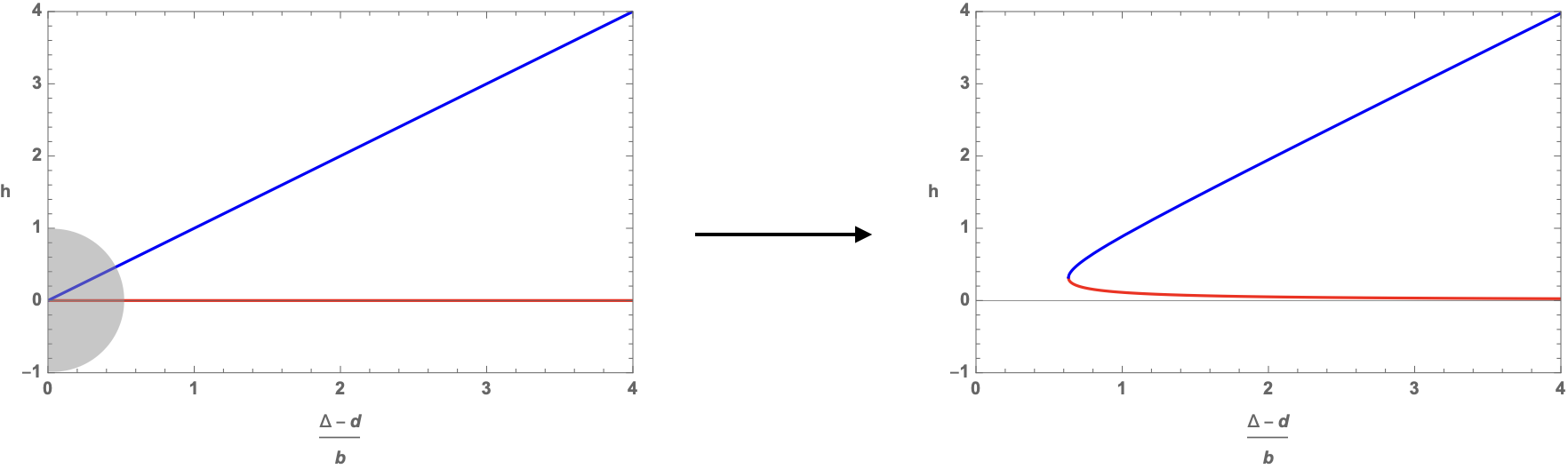}
    \caption{On the left, neglecting $c$ we have the QCD fixed point (orange) crossing with the QCD$_3^*$ fixed point (blue). The grey shaded area represents the region where $\Delta\sim d$ where higher corrections are needed. Upon including $c$ we find the situation on the right, which shows fixed point annihilation.}
    \label{fig:fixed_point_annhiliation}
\end{figure}
It is tempting to conjecture that this scenario applies to both the blue and red regions in Figure  \ref{fig:phase_diagram_fermionicQCD}. The difference is however in the nature of the quartic operator which is hitting marginality. Above the dashed curve, for $\frac{2k}{N_f}>0.78$, the dangerously irrelevant operator is $\mathcal{O}_1=(\psi_{\text{sing}}^2)^2$. Thus, one may perform a Hubbard-Stratonovich transformation in eq. \eqref{eq:SQCD*} and write
\begin{equation}
S_{{\rm QCD}_3^*}\sim S_{{\rm QCD}_3}+h\,\int \sigma_s\,\psi_{\text{sing}}^2 + \cdots \,.
\end{equation}
Clearly, $\sigma_s\sim \psi^2_{\text{sing}}$ is a singlet under the flavor symmetry. Thus, if it were to take a VEV, conformal symmetry would be spontaneously broken but the $\mathrm{U}(N_f)$ symmetry would remain intact.

On the other hand, below the dashed curve, for $\frac{2k}{N_f}<0.78$, the dangerously irrelevant operator is $\mathcal{O}_2=(\psi^2_{\text{adj}})^2$. Thus in this case 
\begin{equation}
S_{{\rm QCD}_3^*}\sim S_{{\rm QCD}_3}+h\,\int \sigma_a\,\psi^2_{\text{adj}} + \cdots \,.
\end{equation}
In this case $\sigma_a\sim \psi^2_{\text{adj}}$ is an adjoint under the flavor symmetry, and thus, if it took a VEV, it would break the global symmetry. From the the Vafa-Witten theorem, as well as from the expectations coming from bosonization dualities \cite{Komargodski:2017keh}, such breaking must be
\begin{align}\label{eq:chiral_sym_breaking_pattern}
    \mathrm{U}\left( N_f \right)\to \mathrm{U}\left(\frac{N_f}{2}+k\right) \times \mathrm{U}\left(\frac{N_f}{2}-k\right)\,.
\end{align}
This is consistent with the fact that when branching the adjoint of $\mathrm{U}(N_f)$ into representations of $\mathrm{U}\left(\frac{N_f}{2}+k\right) \times \mathrm{U}\left(\frac{N_f}{2}-k\right)$, such branching always contains a singlet.

\subsubsection*{Generating scales}

It is natural to wonder how the relevant VEV's for $\sigma_{s,a}$ may come around and what is the mechanism setting their scale. The key observation is that, on general grounds, due to fixed point annihilation the theory gains a mass scale $\Lambda_{IR}$ resulting in loss of conformality \cite{Kaplan:2010zz}. Indeed, from \eqref{eq:betalambdaUNFOLDED} the fixed point merging happens at
\begin{equation}
    \alpha_{\star}=b\,c\,,
\end{equation}
where we have introduced the variable $\alpha=\frac{(d-\Delta)^2}{4}$. Note that, since $\Delta,\,b,\,c$ are generically functions of $N_f$, $N_c$ and $k$, the above equation can be regarded as defining a critical $N_f^{\star}$ for a given $N_c,\,k$. At this point, the coupling takes the value $h_{\star}=\frac{d-\Delta}{2b}$. Writing $h=h_{\star}+\frac{1}{b}\,\epsilon$, the $\beta$ function for small $\epsilon$ is
\begin{equation}
    \beta_{\epsilon}=(\alpha-\alpha_{\star})-\epsilon^2\,,
\end{equation}
Thus, if $\alpha>\alpha_{\star}$ we find two fixed points at $\epsilon=\pm \sqrt{\alpha-\alpha_{\star}}$ which approach each other as one takes $\alpha\rightarrow \alpha_{\star}$ and cease to exist beyond that. Let us now suppose that we start at a scale $\Lambda_{UV}$ where the coupling takes a value $\epsilon_{UV}>0$ and we evolve towards a scale $\Lambda_{IR}$ where the coupling takes a value $\epsilon_{IR}<0$ in the regime where $\alpha$ is slightly smaller than $\alpha_{\star}$. From the definition of the $\beta$ function it follows that
\begin{equation}
    \frac{d\epsilon}{dt}=\beta_{\epsilon}\qquad\leadsto \qquad t_{IR}-t_{UV}=\int_{\epsilon_{UV}}^{\epsilon_{IR}}\frac{d\epsilon}{\beta_{\epsilon}}\,,
\end{equation}
where $t_{UV/IR}=\log\frac{\Lambda_0}{\Lambda_{UV/IR}}$. Hence,
\begin{equation}
    \frac{\Lambda_{IR}}{\Lambda_{UV}}={\rm exp}\Big(-\int_{\epsilon_{UV}}^{\epsilon_{IR}}\frac{d\epsilon}{\beta_{\epsilon}}\Big)\sim e^{-\frac{\pi}{\sqrt{\alpha-\alpha_{\star}}}}\,.
\end{equation}
Thus, an IR scale $\Lambda_{IR}$ exponentially separated from the UV is naturally generated.


\subsubsection{The resulting phase diagram for CS-QCD$_3$}

Putting all ingredients together, our analysis we suggests that the phase diagram is qualitatively as in \ref{fig:ph_diagV4_Annotated} below. 

\begin{figure}[h!]
    \centering
    \includegraphics[width=0.5\textwidth]{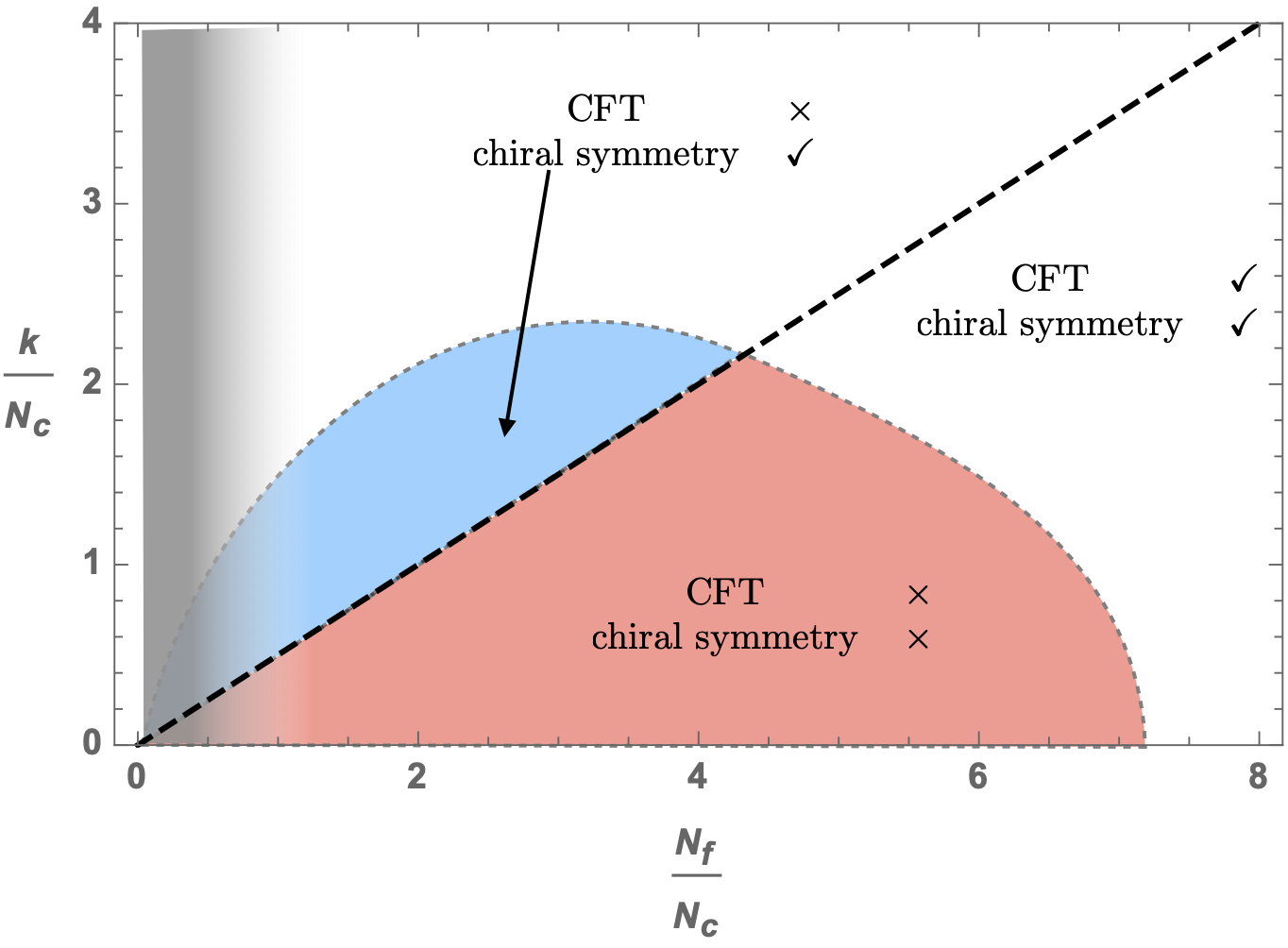}
    \caption{The proposed phase diagram for QCD$_3$ in the plane $\frac{N_f}{N_c}$ and $\frac{k}{N_c}$.}
    \label{fig:ph_diagV4_Annotated}
\end{figure}

Concentrating firstly on the $k\lsim N_f/2$ region of our resulting phase diagram, we see that it is in qualitative agreement with the proposal in \cite{Komargodski:2017keh}: for some critical $N_f^*$ which depends on $N_c$, the flavor symmetry is broken as dictated by the Vafa-Witten theorem. Moreover, the conformal symmetry would be broken by the VEV of the Hubbard-Stratonovich field whose scale is set by $\Lambda_{IR}$. Note that, in order to break the global symmetry as $ \mathrm{U}\left( N_f \right)\to \mathrm{U}\left(\frac{N_f}{2}+k\right) \times \mathrm{U}\left(\frac{N_f}{2}-k\right)$, the Hubbard-Stratonovich field $\sigma_a$, as a traceless $N_f \times N_f$ matrix,  must have $\frac{N_f}{2}+k$ negative equal eigenvalues and $\frac{N_f}{2}-k$ positive equal eigenvalues. Hence $\frac{N_f}{2}+k$ fermionic flavors get a negative mass and $\frac{N_f}{2}+k$ fermionic flavors get a positive mass, so the IR Chern-Simon level $k$ is shifted as
\begin{equation}\label{eq:CS_shift}
   k \quad \rightarrow \quad  k +\frac{1}{2} \left(\frac{N_f}{2}+k\right)(-1) + \frac{1}{2}\left(\frac{N_f}{2}-k\right) = 0 \,,
\end{equation}
so at low energies the TQFT is $\mathrm{SU}(N)_0$, which is trivial.\footnote{For $k\neq 0$ there is also a different logical possibility: that $\sigma_a$ has $\frac{N_f}{2}+k$ positive equal eigenvalues and $\frac{N_f}{2}-k$ negative equal eigenvalues, which would lead to an $\mathrm{SU}(N)_{2k}$ TQFT at low energies.} This is consistent with the proposal of \cite{Komargodski:2017keh}, which at zero mass and low energies flows to a NLSM, with a trivial TQFT. The line separating the regions of flavor symmetry breaking/not breaking is at $\frac{2k}{N_f}\sim 0.78$, slightly below although in reasonable agreement with the expected $\frac{2k}{N_f}=1$.

From our computation, we can extract the critical value $N_f^*$ at which the new phase with broken chiral symmetry opens up. It is simply given by the value of $N_f$ at which $(\psi^2_{\text{adj}})^2$ hits marginality. At $k=0$ the critical value is 
\begin{align}\label{eq:Nf_critial_order1}
    \frac{N_f}{N_c}\Big|_{k=0} = \frac{128}{3\pi^2} \simeq 4.32\,.
\end{align}
This is in exact agreement with the old results in the literature (see \textit{e.g.} \cite{Appelquist:1989tc})\footnote{This agreement may not be completely unexpected, as, in our regime, the quartic operators in question are computed \emph{via} fermion bilinears. This is precisely the same quantity which \cite{Appelquist:1989tc} considers.} and in qualitatively agreement with the $\epsilon$ expansion results in \cite{Goldman:2016wwk}, which finds $\frac{N_f}{N_c}=\frac{11}{2}+\mathcal{O}(\frac{1}{N_c})\sim 5.5$. 

It is important to stress that while the critical value for $N_f^*$ \eqref{eq:Nf_critial_order1} doesn't land squarely outside the regime of validity of our approximation (which, recall, strictly speaking, demands $\frac{N_f}{N_c}\gg 1$); it is such that one may worry about the legitimacy of the leading order approximation. A possible way to strengthen the validity of our results would be to compute the anomalous dimensions at higher order in $1/N_f$. However, this computation, for any $k$ and $N_c$, is very involved, and we have not carried it out. Still, for $k=0$, the result at order $1/N_f^2$ is already known \cite{Gracey:2018fwq},
\begin{eqnarray}\label{eq:anomalousdim_order2}
    \Delta(\psi^2_{\text{adj}}) &=& 2-\frac{64}{3\pi^2}\,\frac{N_c^2-1}{N_c\,N_f}-\frac{256\,(N_c^2-1)\,\Big(2N_c^2\,(2\pi^2-5)+28-3\pi^2\Big)}{9\pi^4\,N_c^2\,N_f^2}+\cdots\\ && \sim 2-\frac{64}{3\pi^2}\,\frac{N_c}{N_f}-\frac{512\,(2\pi^2-5)\,N_c^2}{9\pi^4\,N_f^2}+\cdots\,.
\end{eqnarray}
We can use this result and compare it with \eqref{eq:Nf_critial_order1}, in order to, at least, reassure as in respect of the fate of the merging of fixed points at $k=0$. From \eqref{eq:anomalousdim_order2}, the critical value of $N_f$ that we find for the phase transition is
\begin{align}\label{eq:Nf_critial_order2}
    \frac{N_f}{N_c}\Big|_{k=0} = \frac{32(\sqrt{2\pi^2-1}+2)}{3\pi^2}\simeq 6.84\,.
\end{align}
Importantly, while the difference between \eqref{eq:Nf_critial_order1} and \eqref{eq:Nf_critial_order2} is significant, it is such that the value of $N_f^*$ grows, not decreases. This is evidence that the two regions of Figure \ref{fig:phase_diagram_fermionicQCD} where the fixed points merge do survive after taking into account higher $1/N_f$ corrections. One can go one step further, and, having the first two terms of the $1/N_f$ expansion, perform a Padé resumation to try and get an estimate for the full result. There are two ways to find the Padé approximant, which lead to values of $N_f^*/N_c$ in the range of 6-8.\footnote{The two possible Pade resummations of \eqref{eq:anomalousdim_order2} are
\begin{equation}
\eqref{eq:anomalousdim_order2}\sim\frac{2+a_1\,\frac{N_c}{N_f}}{1+a_2\,\frac{N_c}{N_f}}=\frac{2+\frac{16-32\pi^2}{3\pi^2}\,\frac{N_c}{N_f}}{1+\frac{40-16\pi^2}{3\pi^2}\,\frac{N_c}{N_f}}\,,\qquad {\rm or}\qquad \eqref{eq:anomalousdim_order2}\sim\frac{2}{1+b_1\,\frac{N_c}{N_f}+b_2\,\frac{N_c^2}{N_f^2}}=\frac{2}{1+\frac{32}{3\pi^2}\,\frac{N_c}{N_f}-\frac{256-512\pi^2}{9\pi^4}\,\frac{N_c^2}{N_f^2}}\,.
\end{equation}
where the $a_i$, $b_i$ coefficients are fixed expanding for small $\frac{N_c}{N_f}$ and fitting to \eqref{eq:anomalousdim_order2}. Then, the estimation of the critical $\frac{N_f}{N_c}$ gives  $\frac{N_f}{N_c}\sim 8.3$ and $\frac{N_f}{N_c}\sim 6$, respectively.} Basically, the second order correction in \eqref{eq:anomalousdim_order2} turns out to be negative and small, so the first order result is confirmed.

In turn, in the region $k\gsim N_f/2$, our proposed phase diagram in Figure \ref{fig:ph_diagV4_Annotated} contains a blue region with broken conformal symmetry and unbroken flavor symmetry, arising from the VEV of the Hubbard-Stratonovich flavor singlet. The existence of this region is consistent with \cite{Armoni:2019lgb}. 

Our scenario suggests implications for the shape of the red curve, where the phase transition from the conformal phase to the non-conformal one takes place. It was pointed out in \cite{Komargodski:2017keh} that there is a constraint on the values of $N_f^*$, coming from the consistency of mass-deforming the theory in the UV and the IR,
\begin{align}
    N_f^*(k) - N_f^*(k\pm \frac{1}{2}) \le 1\,.
\end{align}
This can be rewritten as
\begin{align}\label{eq:constraint_Nf}
    -1\le N_f^*(k) - N_f^*(k-\frac{1}{2})\le 1\,,
\end{align}
which is essentially a constraint on the slope of the red curve in Figure \ref{fig:ph_diagV4_Annotated}. On the other hand, our first order calculation shows that for growing Chern-Simons level $k$, the critical value of $N_f^*$ decreases.\footnote{One may worry that this is an artefact of considering only the leading approximation in $\frac{1}{N_f}$. A plausibility argument comes from considering the lagrangian of CS-QCD with the normalization of the gauge field such that the coupling constant appears only in the interaction term (and not in the denominator of the $F^2$ term). Then, upon redefining $A$ to make the level $k$ disappear from the numerator in the Chern-Simons term; it reappears in the denominator of the interaction term, as $\propto g/\sqrt{k}$. Therefore, increasing $k$ at fixed $N_c$ and $N_f$ will effectively result in a smaller gauge coupling, and with it smaller anomalous dimensions, which in turn suggests that the merging of fixed points will occur for smaller values of $N_f$.} Accordingly, the prediction, from the point of view of merging of fixed points, would be that the critical value of the number of fermions is subject to a stronger constraint than \eqref{eq:constraint_Nf}, namely
\begin{align}
    -1\le N_f^*(k) - N_f^*(k-\frac{1}{2})\le 0\,.
\end{align}

\subsubsection{Domain walls of QCD$_4$: a relation between 3d and 4d conformal windows?}

Before concluding, let us comment on the relation between the theories we study and domain walls in 4d. In \cite{Gaiotto:2017tne}, it was proposed that $4$d QCD with gauge group $\mathrm{SU}(N)$ and $N_f$ Dirac fermionic flavors, at $\theta= \pi$, has two vacua which can be separated by a domain wall described by $3$d $\mathrm{SU}(N)_{1-N_f/2}$ with $N_f$ fermionic flavors. Notice that the $4$d fermions are $4$ component fermions, while the $3$d fermions are $2$ component fermions.

The $3$d domain wall theory $\mathrm{SU}(N)_{1-N_f/2}$ with $N_f$ flavors lies almost at the boundary of the region $|k| \leq N_f/2$. At large positive masses, it flows to the TQFT $\mathrm{SU}(N)_1 \leftrightarrow \mathrm{U}(1)_{N}$ (using level-rank duality), while at large negative masses it flows to $\mathrm{SU}(N)_{1+N_f} \leftrightarrow \mathrm{U}(N_f+1)_{-N}$. In between these two regimes, there should be a quantum phase described by a NLSM with target space $\mathbb{CP}^{N_f-1}$. The expectation is that $N_f > (N_f^*)_{3d}$ the $3d$  domain wall theory flows to a CFT with $\mathrm{U}(N_f)$ symmetry, while if $N_f < (N_f^*)_{3d}$ the $3d$  domain wall theory flows to a NLSM with target space $\mathbb{CP}^{N_f-1}$.

From results in this paper, valid at large $N_f \gg N_c \gg 1$, when taken to the lower bound of their validity region, we expect $(N_f^*)_{3d}$ for the domain wall theories to be approximately half (but because of the constraint on the slope of $N_f*(k)$ it must be a bit larger than) of $(N_f^*)_{3d}$ at $k=0$, which at second order in $1/N_f$ is about $(6 \sim 8) N_c$. So we expect that for the $3$d domain wall theories 
\begin{equation}
\frac{(N_f^*)_{3d}}{N_c} \simeq 3 \sim 4\,. 
\end{equation}
This number turns out to be close to the lower bound of the conformal window of QCD$_4$.

Thus, $(N_f^*)_{3d}$ is tantalizingly similar to $(N_f^*)_{4d}$, i.e. the lower bound of the conformal window in 4d QCD. In other words, it looks like that the $4$d bulk theory is conformal and preserves the full symmetry if and only if the $3$d domain wall theory does. The possibility of this connection was already alluded to in \cite{Gaiotto:2018yjh}. 

A possible physical mechanism which explains the equality between $(N_f^*)_{3d}$ and $(N_f^*)_{4d}$ might be that the QCD$^*_3$ considered here, that is, $\mathrm{SU}(N)_{1-N_f/2}$ with $N_f$ fermions and quartic interactions $(\psi^2_{adj})^2$ turned on; describes the domain walls of QCD$^*_4$, that is, the hypotethical $4$d theory which is supposed to merge with QCD$_4$ at the lower edge of the $4$d conformal window \cite{Kaplan:2010zz}. We believe that this observation deserves further qualitative and quantitative investigation.

\subsection{Flavor non-singlet quartic operators}\label{sec:non_singlets}
We previously argued that, in the large $N_f$ limit, QCD$_3$ has de-stabilizing dangerously irrelevant operators due to a nearby fixed point --dubbed QCD$_3^*$-- with which eventually collides and annihilates, prompting the breaking of conformal invariance as well as of the flavor symmetry in a pattern consistent with the proposed phase diagram for QCD$_3$. It is interesting to study other quartic operators in non-trivial representations under the flavor symmetry. In the following we will restrict to $k=0$ and focus on the operators transforming in the $[2,0,0,\dots,0,0,2]$ and $[0,1,0,\dots,0,1,0]$ of the $\mathrm{SU}(N_f)\subseteq \mathrm{U}(N_f)$ flavour symmetry. There are four different operators, two in each representation, corresponding to the single and double trace versions with respect to the contractions of the colour indices. The upshot of the discussion is that these operators hit marginality at precisely the same point as the singlets discussed in the main text. Moreover, the two representations that they transform in are such that their branching rule under the symmetry breaking pattern \eqref{eq:chiral_sym_breaking_pattern} will give rise to a singlet of $\mathrm{U}\left(\frac{N_f}{2}\right) \times \mathrm{U}\left(\frac{N_f}{2}\right)$. It is an alluring coincidence that, at the same time as the mechanism of fixed point merging gives rise to the breaking of chiral symmetry, a new host of operators, singlets under the remaining symmetry, become relevant and available to be added to the lagrangian.

There are two possible ways to proceed with the computation. One is to first identify a basis of the quartic operators that explicitly separates them in the aforementioned representations, in which case we will not encounter any mixing among them. Another is to choose four independent quartic operators and untangle their mixing later. We opt for the second option.

We use the following basis for the operators. The double trace operators are
\begin{align}
O_1^d=\left(\bar{\psi}_{1a}\psi^{3a}\right)\left(\bar{\psi}_{2b}\psi^{4b}\right)\\
O_{2}^d=\left(\bar{\psi}_{1a}\psi^{4a}\right)\left(\bar{\psi}_{2b}\psi^{3b}\right)
\end{align}
where $a,b$ are colour indices in the fundamental of $\mathrm{SU}(N_c)$, 1,2,3,4 are flavour indices and the parenthesis indicate contraction of the spinor indices. On the other hand, the single trace operators are
\begin{align}
O_1^s=\left(\bar{\psi}_{1a}\psi^{3b}\right)\left(\bar{\psi}_{2b}\psi^{4a}\right)\\
O_2^s=\left(\bar{\psi}_{1a}\psi^{4b}\right)\left(\bar{\psi}_{2b}\psi^{3a}\right)
\end{align}

To compute the dimensions of the operators we will follow the same method as for the singlets, namely, identifying the coefficient of the logarithmic divergence. We compile the relevant technical details in appendix \ref{app:computation_non_singlets}. We now have the extra complication of operator mixing. At zero-th order in the large $N_f$ expansion, we can write the matrix of correlators as 
\begin{align}
\langle O_\alpha(x)\bar{O}_\beta(0)\rangle_{(0)} = \frac{\mathcal{C}^{(0)}_{\alpha\beta}}{x^8}
\end{align}
Ordering the basis as $\lbrace O_1^d,O_2^d,O_1^s,O_2^s  \rbrace$, the matrix $\mathcal{C}^{(0)}$ is 
\begin{align}
\mathcal{C}^{(0)}=\frac{1}{128\pi^4}\left(
\begin{array}{cccc}
2N_c^2&-N_c&2N_c&-N_c^2\\
-N_c&2N_c^2&-N_c^2&2N_c\\
2N_c&-N_c^2&2N_c^2&-N_c\\
-N_c^2&2N_c&-N_c&2N_c^2
\end{array}\right)
\end{align}
At first order in $1/N_f$, we know we can write the matrix of correlators in the form
\begin{align}
\langle O_\alpha(x)\bar{O}_\beta(0)\rangle_{(1)}=-\frac{\mathcal{C}^{(1)}_{\alpha\beta}}{x^8}\log\left(x^2\Lambda^2\right) + \text{finite}
\end{align}
With the same choice of basis as before, the matrix $\mathcal{C}^{(1)}$ is
\begin{align}
\mathcal{C}^{(1)}=\left(
\begin{array}{cccc}
-\frac{2N_c\left(N_c^2-1\right)}{3\pi^6N_f}&\frac{5\left(N_c^2-1\right)}{6\pi^6N_f}&-\frac{2\left(N_c^2-1\right)}{3\pi^6N_f}&\frac{N_c\left(N_c^2-1\right)}{3\pi^6N_f}\\
\frac{5\left(N_c^2-1\right)}{6\pi^6N_f}&-\frac{2N_c\left(N_c^2-1\right)}{3\pi^6N_f}&\frac{N_c\left(N_c^2-1\right)}{3\pi^6N_f}&-\frac{2\left(N_c^2-1\right)}{3\pi^6N_f}\\
-\frac{2\left(N_c^2-1\right)}{3\pi^6N_f}&\frac{N_c\left(N_c^2-1\right)}{3\pi^6N_f}&-\frac{N_c\left(N_c^2-1\right)}{6\pi^6N_f}&\frac{5\left(N_c^2-1\right)}{6\pi^6N_f}\\
\frac{N_c\left(N_c^2-1\right)}{3\pi^6N_f}&-\frac{2\left(N_c^2-1\right)}{3\pi^6N_f}&\frac{5\left(N_c^2-1\right)}{6\pi^6N_f}&-\frac{N_c\left(N_c^2-1\right)}{6\pi^6N_f}
\end{array}\right)
\end{align}
As in \cite{Chester:2016ref}, the anomalous dimensions are the eigenvalues of $(\mathcal{C}^{(0)})^{-1}\mathcal{C}^{(1)}$.
\begin{align}
(\mathcal{C}^{(0)})^{-1}\mathcal{C}^{(1)}=
\left(
\begin{array}{cccc}
-\frac{64 \left(2 N_c^2-1\right)}{3 \pi ^2 N_c N_f} & \frac{128}{3 \pi ^2 N_f} & -\frac{64}{3 \pi ^2 N_f} & \frac{64
	\left(N_c^2-2\right)}{3 \pi ^2 N_c N_f} \\
\frac{128}{3 \pi ^2 N_f} & -\frac{64 \left(2 N_c^2-1\right)}{3 \pi ^2 N_c N_f} & \frac{64 \left(N_c^2-2\right)}{3 \pi ^2 N_c
	N_f} & -\frac{64}{3 \pi ^2 N_f} \\
\frac{64}{3 \pi ^2 N_f} & -\frac{128}{3 \pi ^2 N_c N_f} & \frac{64}{3 \pi ^2 N_c N_f} & \frac{64}{3 \pi ^2 N_f} \\
-\frac{128}{3 \pi ^2 N_c N_f} & \frac{64}{3 \pi ^2 N_f} & \frac{64}{3 \pi ^2 N_f} & \frac{64}{3 \pi ^2 N_c N_f} \\
\end{array}
\right)
\end{align}
This matrix is diagonal in the following basis
\begin{align}
A_1&=\frac{2N_c^2+N_c+\sqrt{4N_c^4-11N_c^2+16}}{2(N_c+2)} \left(O_1^d-O_2^d\right)-\left(O_1^s-O_2^s\right)\\
S_1&=-\frac{2N_c^2-N_c+\sqrt{4N_c^4-11N_c^2+16}}{2(N_c-2)} \left(O_1^d+O_2^d\right) + \left(O_1^s+O_2^s\right)\\
A_2&=\frac{2N_c^2+N_c-\sqrt{4N_c^4-11N_c^2+16}}{2(N_c+2)}\left(O_1^d-O_2^d\right)-\left(O_1^s-O_2^s\right)\\
S_2&=-\frac{2N_c^2-N_c-\sqrt{4N_c^4-11N_c^2+16}}{2(N_c-2)}\left(O_1^d+O_2^d\right)+\left(O_1^s+O_2^s\right)
\end{align}
and the anomalous dimensions are
\begin{align}
\Delta_{A_1}=4-\frac{32\left(N_c(2N_c+3)-2+\sqrt{4N_c^4-11N_c^2+16}\right)}{3\pi^2N_c N_f}\\
\Delta_{S_1}=4-\frac{32\left(N_c(2N_c-3)-2+\sqrt{4N_c^4-11N_c^2+16}\right)}{3\pi^2N_c N_f}\\
\Delta_{A_2}=4-\frac{32\left(N_c(2N_c+3)-2-\sqrt{4N_c^4-11N_c^2+16}\right)}{3\pi^2N_c N_f}\\
\Delta_{S_2}=4+\frac{32\left(N_c(3-2N_c)+2+\sqrt{4N_c^4-11N_c^2+16}\right)}{3\pi^2N_c N_f}
\end{align}

At large $N_c$, the first two of these operators hit marginality at precisely \eqref{eq:Nf_critial_order1}.

\section{Bosonic CS-QCD}\label{sec:bosonic}

 \begin{figure}[b]
	\centering
	\begin{tikzpicture} [scale=0.5]  
	\draw[dashed, mygreen, thick] (-0.8,0) to (1.2,0);
	\node at (2.1,-0.05) {$=$};
	\draw[dashed,thick] (3,0) to (5,0);
	\node at (5.5,0) {$+$};
	\draw[dashed,thick] (6,0) to (7.1,0);
	\draw[thick, greenZS] (8.1,0) circle(1cm);
	\draw[dashed,thick] (9.1,0) to (10.2,0);
	\node at (10.7,0) {$+$};
	\draw[dashed,thick] (11.3,0) to (12.4,0);
	\draw[thick, greenZS] (13.4,0) circle(1cm);
	\draw[dashed, thick] (14.4,0) to (15.5,0);
	\draw[thick,greenZS] (16.5,0) circle (1cm);
	\draw[dashed,thick] (17.5,0) to (18.6,0);
	\node at (19,0) {$ \ \ \  \ \ \ + \  \cdots$};     
	\end{tikzpicture}
	\caption{(bQCD) HS field $\sigma_s$ effective propagator (green dashed line). The black dashed line stands for tree level HS field propagator. The diagrams for the $\sigma_A$ field are analogous.} \label{HSresum}
\end{figure}
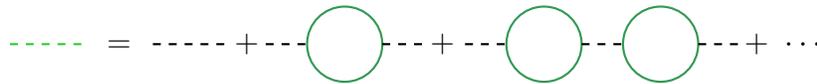

In this section, we perform an analogous analysis of the possible phases for bosonic QCD in 3 dimensions. More precisely, we consider $\mathrm{U}(N_c)_k$ gauge theory coupled to $N_f$ complex scalar fields $\phi_i$. The starting point is the lagrangian
\begin{align}
    \mathcal{L} = \frac{1}{4g^2} F^2 + \sum_i |D_\mu\phi_i|^2 + V(\phi)\,,
\end{align}
The quartic term in the scalar potential is classically relevant, and we can write it as follows,
\begin{align}\label{eq:potential_bosonic}
    V(\phi) = \lambda_s\,  \left( \phi^*_{ia}\phi^{ia}\right)^2 + \lambda_A \,\phi^*_{ia}\phi^{ib}\phi^*_{jb}\phi^{ja}\, ,
\end{align}
As before, $i,j$ are color indices while $a,b$ are flavor. Note that each therm is the square, respectively, of a flavor singlet $\phi_{\text{sing}}=\phi^*_{ia}\phi^{ia}$ and a flavor adjoint $\phi_{\text{adj}}=\phi^*_{ia}\phi^{ib}$.

The first step is to perform a Hubbard-Stratonovich transformation and introduce two auxiliary scalar fields $\sigma_s$ and $\sigma_A$, such that integrating them out will lead to $\sigma_s\sim \lambda_s \phi^*_{ia}\phi^{ia}$ and $\sigma_A\sim \lambda_A \phi^*_{ia}\phi^{ib}$. This results in the following potential,
\begin{align}
    V(\phi,\sigma_s,\sigma_A) = \sigma_s\sum_i \phi^*_{ia}\phi^{ia} + \sigma_A(T^A)^a_b\sum_i\phi^*_{ia}\phi^{ib}\, ,
\end{align}
where $A$ is an index in the adjoint of $\mathrm{SU}(N_c)$. In the transformation, quadratic terms will appear for the HS fields; however, since their classical scaling dimension is equal to 2, they are irrelevant.

\begin{figure}
	\centering
	\begin{tikzpicture}[scale=0.4]
	
	\draw[black] (-4,3) to (22,3);
	\draw[black] (-4,3) to (-4,-30);
	\draw[black] (22,3) to (22,-30);
	\draw[black] (-4,-30) to (22,-30);
	
	\draw[gluon,red] (-2,0) to (2,0);
	\node at (-2,0.8) {\scriptsize \textcolor{red}{$A$}};
	\node at (2,0.8) {\scriptsize \textcolor{red}{$B$}};
	\node at (-2,-0.8) {\scriptsize \textcolor{red}{$\mu$}};
	\node at (2,-0.8) {\scriptsize \textcolor{red}{$\nu$}};
	\node at (12,0) {$=\frac{16 \, (N_c \delta^{AB}+1)}{N_f N_c (1+\lambda^2)|p|}\left(\delta_{\mu\nu}-\xi \frac{p_\mu p_\nu}{p^2}-\lambda \frac{p^\alpha}{|p|}\epsilon_{\alpha\mu\nu}\right)$};

	\draw[fermion] (-2,-4) to (2,-4);
	\node at (-2,-3.5) {\scriptsize\textcolor{greenZS}{$a$}};
	\node at (2,-3.5) {\scriptsize\textcolor{greenZS}{$b$}};
	\node at (-2,-4.5) {\scriptsize\textcolor{greenZS}{$i$}};
	\node at (2,-4.5) {\scriptsize\textcolor{greenZS}{$j$}};
	\node at (5.1,-4) {$=\frac{i}{p^2}\delta_a^b\delta_i^j$};
	
    \draw[dashed,red] (-2,-8) to (2,-8);
    \node at (-2,-7.5) {\scriptsize\textcolor{red}{$A$}};
    \node at (2,-7.5) {\scriptsize\textcolor{red}{$B$}};
    \node at (5.5,-8) {$=\frac{16\, \delta^{AB}}{N_f}|p|$};
    
    \draw[dashed,greenZS] (-2,-11) to (2,-11);
    \node at (5.5,-11) {$=\frac{16}{N_f N_c}|p|$};

    \draw[gluon,red] (-2,-11-4) to (0.5,-15);
	\draw[fermion] (2,-13.5) to (0.5,-15);
	\draw[fermion] (0.5,-15) to (2,-16.5);
	\node at (-2,-14.2) {\scriptsize\textcolor{red}{$A$}};
	\node at (-2,-15.8) {\scriptsize\textcolor{red}{$\mu$}};
	\node at (1.25,-13.35) {\scriptsize\textcolor{greenZS}{$a$}};
	\node at (2.45,-13.5) {\scriptsize\textcolor{greenZS}{$i$}};
	\node at (1.25,-16.5) {\scriptsize\textcolor{greenZS}{$b$}};
	\node at (2.35,-16.5) {\scriptsize\textcolor{greenZS}{$j$}};
	\node at (1.45,-14.6) {\scriptsize\textcolor{greenZS}{$p$}};
	\node at (1.35,-15.3) {\scriptsize\textcolor{greenZS}{$q$}};
	\node at (9.2,-15) {$= i\, \delta^j_i \left[ (T^A)^b_a + \delta^b_a \right] (p+q)^\mu$};

	\draw[dashed,red] (-2,-11-4-4) to (0.5,-15-4);
	\draw[fermion] (2,-13.5-4) to (0.5,-15-4);
	\draw[fermion] (0.5,-15-4) to (2,-16.5-4);
	\node at (-2,-18.5) {\scriptsize\textcolor{red}{$A$}};
	\node at (1.25,-13.35-4) {\scriptsize\textcolor{greenZS}{$a$}};
	\node at (2.45,-13.5-4) {\scriptsize\textcolor{greenZS}{$i$}};
	\node at (1.25,-16.5-4) {\scriptsize\textcolor{greenZS}{$b$}};
	\node at (2.35,-16.5-4) {\scriptsize\textcolor{greenZS}{$j$}};
	\node at (6,-19) {$=i\, (T^A)^b_a\,\delta^j_i$};

	\draw[dashed,greenZS] (-2,-11-4-8) to (0.5,-15-8);
	\draw[fermion] (2,-13.5-8) to (0.5,-15-8);
	\draw[fermion] (0.5,-15-8) to (2,-16.5-8);
	\node at (1.25,-13.35-8) {\scriptsize\textcolor{greenZS}{$a$}};
	\node at (2.45,-13.5-8) {\scriptsize\textcolor{greenZS}{$i$}};
	\node at (1.25,-16.5-8) {\scriptsize\textcolor{greenZS}{$b$}};
	\node at (2.35,-16.5-8) {\scriptsize\textcolor{greenZS}{$j$}};
	\node at (5.4,-19-4) {$=i\, \delta^b_a\,\delta^j_i$};
	
	\draw[gluon,red] (-2,-13.5-12) to (-0,-15-12);
	\draw[gluon,red] (-0,-15-12) to (-2,-16.5-12);
	\draw[fermion] (2,-13.5-12) to (0,-15-12);
	\draw[fermion] (0,-15-12) to (2,-16.5-12);
	\node at (1.25,-13.35-12) {\scriptsize\textcolor{greenZS}{$a$}};
	\node at (2.45,-13.5-12) {\scriptsize\textcolor{greenZS}{$i$}};
	\node at (1.25,-16.5-12) {\scriptsize\textcolor{greenZS}{$b$}};
	\node at (2.35,-16.5-12) {\scriptsize\textcolor{greenZS}{$j$}};
	\node at (-1.25,-13.15-12) {\scriptsize\textcolor{red}{$A$}};
	\node at (-2.35,-13.75-12) {\scriptsize\textcolor{red}{$\mu$}};
	\node at (-1.25,-16.85-12) {\scriptsize\textcolor{red}{$B$}};
	\node at (-2.25,-16.35-12) {\scriptsize\textcolor{red}{$\nu$}};
	\node at (10,-15-12) {$=-\delta^{\mu\nu}\left[ (T^A)^x_a (T^B)_x^b + \delta^b_a\right] \delta^j_i$};

	\end{tikzpicture}
	\caption{Summary of Feynman rules of bosonic QCD, after resumming the fermion bubbles in the gluon propagator. $A$ and $B$ are labels in the adjoint of the gauge group, while $a$ and $b$ label the fundamental representation. The flavour indices are $i,j$; while $\mu,\nu$ are spacetime indices as usual.} \label{fig:bQCDfeynmanrules}
\end{figure}

In order to find the Feynman rules in the large $N_f$ regime, we proceed in a similar way to section \ref{sec:fermionic}, i.e. we sum the fermion bubbles for the HS fields as in Figure \ref{HSresum}. This results in the following effective propagators, 
\begin{align}
&\langle\sigma_s(x)\sigma_s(0)\rangle_{\text{eff}}=\frac{8}{\pi^2N_cN_fx^4}\, ,\\
&\langle\sigma_A(x)\sigma_B(0)\rangle_{\text{eff}}=\frac{8\ \delta_{AB}}{\pi^2N_fx^4}\, .
\end{align}
We also do the same bubble resumation for the gauge fields, and add the non-local gauge fixing term \eqref{eq:nonlocal_gf} as well as the standard Chern-Simons term. The resulting Feynman rules are summarized in Figure \ref{fig:bQCDfeynmanrules}. Note that we have put together the $\mathrm{U}(1)$ and $\mathrm{SU}(N_c)$ parts of the $\mathrm{U}(N_c)$ gauge field, as they will always contribute in the same way.

As in the fermionic case, we are interested in searching for dangerously irrelevant operators. The natural candidates are the irrelevant operators closest to marginality. In this case, the operators in question are $\phi^2_{\text{sing}}$ and $\phi^2_{\text{adj}}$. Note however that the HS transformation that we have performed forces us to substitute $\phi^2_{\text{sing}}$ by the HS field $\sigma_s$. In conclusion, we need to compute the logarithmic divergences of the $\langle \phi^2_{\text{adj}}(x) \phi^2_{\text{adj}}(0)\rangle$ and $\langle \sigma_s(x)\sigma_s(0) \rangle$ correlators at order $1/N_f$. However, just as in the fermionic case, further simplifications arise in the large $N_c$ limit (provded of course $N_f\gg N_c\gg 1$). In this case, owing to large $N_c$ factorization, the dimension of the quartics will be twice that of the bilinears. Thus, our task will be to compute the dimension of $\phi_{\text{adj}}$ and $\sigma_s$ and search for the locus in parameter space where these become $\frac{3}{2}$.

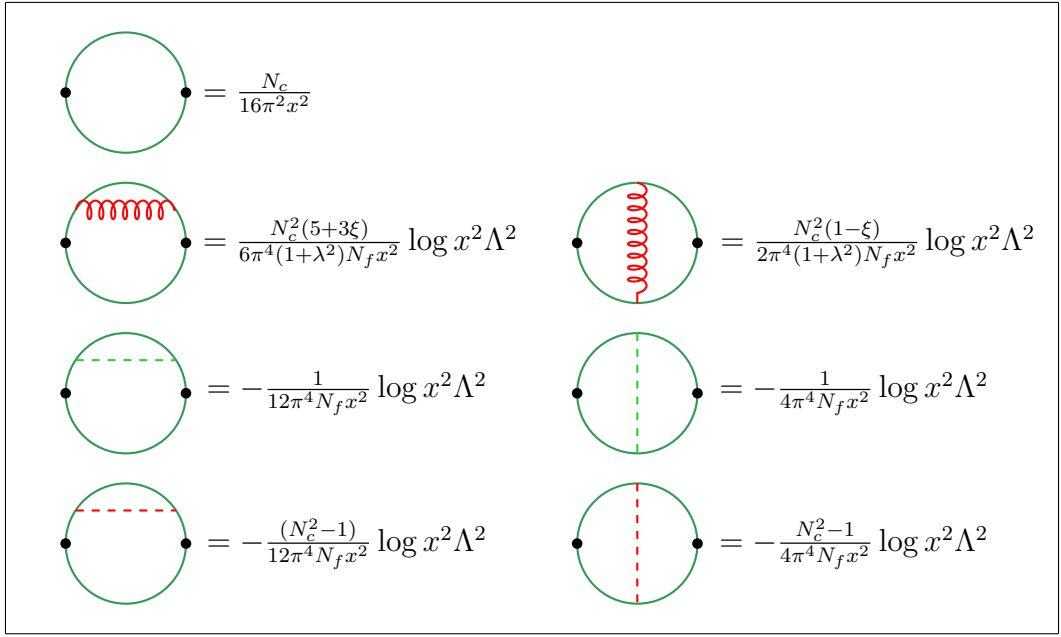
\begin{figure}[h!]
	\centering
	\begin{tikzpicture}[scale=0.4]
	
	\draw[black] (-4-5,3) to (21+5,3);
	\draw[black] (-4-5,3) to (-4-5,-40+2+5+10+5);
	\draw[black] (21+5,3) to (21+5,-40+2+5+10+5);
	\draw[black] (-4-5,-40+2+5+10+5) to (21+5,-40+2+5+10+5);
	
	\draw[thick, greenZS] (0-5,0) circle (2cm);
	\draw[thick,fill] (-2-5,0) circle (4pt);
	\draw[thick,fill] (2-5,0) circle (4pt);
	
	\node at (5-5-0.6,0) {$=\frac{N_c}{16\pi^2x^2}$};
	
	\draw[greenZS,thick] ([shift=(180:2cm)]0-5,-5)   arc (180:0:2cm);
	\draw[greenZS,thick] ([shift=(0:2cm)]0-5,-5)   arc (0:-180:2cm);
	\draw[gluon,red] (-1.66-5,-3.9) to (1.66-5,-3.9);
	\draw[thick,fill] (-2-5,-5) circle (4pt);
	\draw[thick,fill] (2-5,-5) circle (4pt);
	\node at (7.8-5,-5) {$ =\frac{N_c^2(5+3\xi)}{6\pi^4(1+\lambda^2)N_fx^2}\log x^2\Lambda^2$};
	
		\draw[greenZS,thick] ([shift=(180:2cm)]10+2,-5)   arc (180:0:2cm);
	\draw[greenZS,thick] ([shift=(0:2cm)]10+2,-5)   arc (0:-180:2cm);
	\draw[gluon,red] (10+2,-0) ([shift=(90:2cm)]10+2,-5)  to   ([shift=(-90:2cm)]10+2,-5);
	\draw[thick,fill] (8+2,-5) circle (4pt);
	\draw[thick,fill] (12+2,-5) circle (4pt);
	\node at (15+5,-5) {$=\frac{N_c^2(1-\xi)}{2\pi^4(1+\lambda^2)N_fx^2}\log x^2\Lambda^2$};

	\draw[greenZS,thick] ([shift=(180:2cm)]0-5,-10)   arc (180:0:2cm);
	\draw[greenZS,thick] ([shift=(0:2cm)]0-5,-10)   arc (0:-180:2cm);
	\draw[dashed, thick,mygreen] (-1.66-5,-8.9) to (1.66-5,-8.9);
	\draw[thick,fill] (-2-5,-10) circle (4pt);
	\draw[thick,fill] (2-5,-10) circle (4pt);
	\node at (8.05-5-0.7,-10) {$ =-\frac{1}{12\pi^4N_f x^2}\log x^2\Lambda^2$};

	\draw[greenZS,thick] ([shift=(180:2cm)]0+12,-10)   arc (180:0:2cm);
	\draw[greenZS,thick] ([shift=(0:2cm)]0+12,-10)   arc (0:-180:2cm);
	\draw[dashed,thick,mygreen] (0+12,-5) ([shift=(90:2cm)]0+12,-10)  to   ([shift=(-90:2cm)]0+12,-10);
	\draw[thick,fill] (-2+12,-10) circle (4pt);
	\draw[thick,fill] (2+12,-10) circle (4pt);
	\node at (7.9+12-0.7,-10) {$=-\frac{1}{4\pi^4N_fx^2}\log x^2\Lambda^2$};

	\draw[greenZS,thick] ([shift=(180:2cm)]0-5,-15)   arc (180:0:2cm);
	\draw[greenZS,thick] ([shift=(0:2cm)]0-5,-15)   arc (0:-180:2cm);
	\draw[dashed,thick,red] (-1.66-5,-8.9-5) to (1.66-5,-8.9-5);
	\draw[thick,fill] (-2-5,-15) circle (4pt);
	\draw[thick,fill] (2-5,-15) circle (4pt);
	\node at (8.05-5-0.7,-15) {$ =-\frac{(N_c^2-1)}{12\pi^4N_fx^2}\log x^2\Lambda^2$};
		
	\draw[greenZS,thick] ([shift=(180:2cm)]0+12,-15)   arc (180:0:2cm);
	\draw[greenZS,thick] ([shift=(0:2cm)]0+12,-15)   arc (0:-180:2cm);
	\draw[dashed,thick,red] (0+12,-5) ([shift=(90:2cm)]0+12,-15)  to   ([shift=(-90:2cm)]0+12,-15);
	\draw[thick,fill] (-2+12,-15) circle (4pt);
	\draw[thick,fill] (2+12,-15) circle (4pt);
	\node at (7.9+12-0.7,-15) {$=-\frac{N_c^2-1}{4\pi^4N_fx^2}\log x^2\Lambda^2$};

	\end{tikzpicture}
	\caption{ (bQCD) Results for individual Feynman diagrams appearing in the 2-point correlation function of the scalar-bilinear adjoint operators. The green dashed line stands for the effective HS singlet, and the red dashed line for the effective adjoint HS field.} \label{bQCDadjfeynman}
\end{figure}

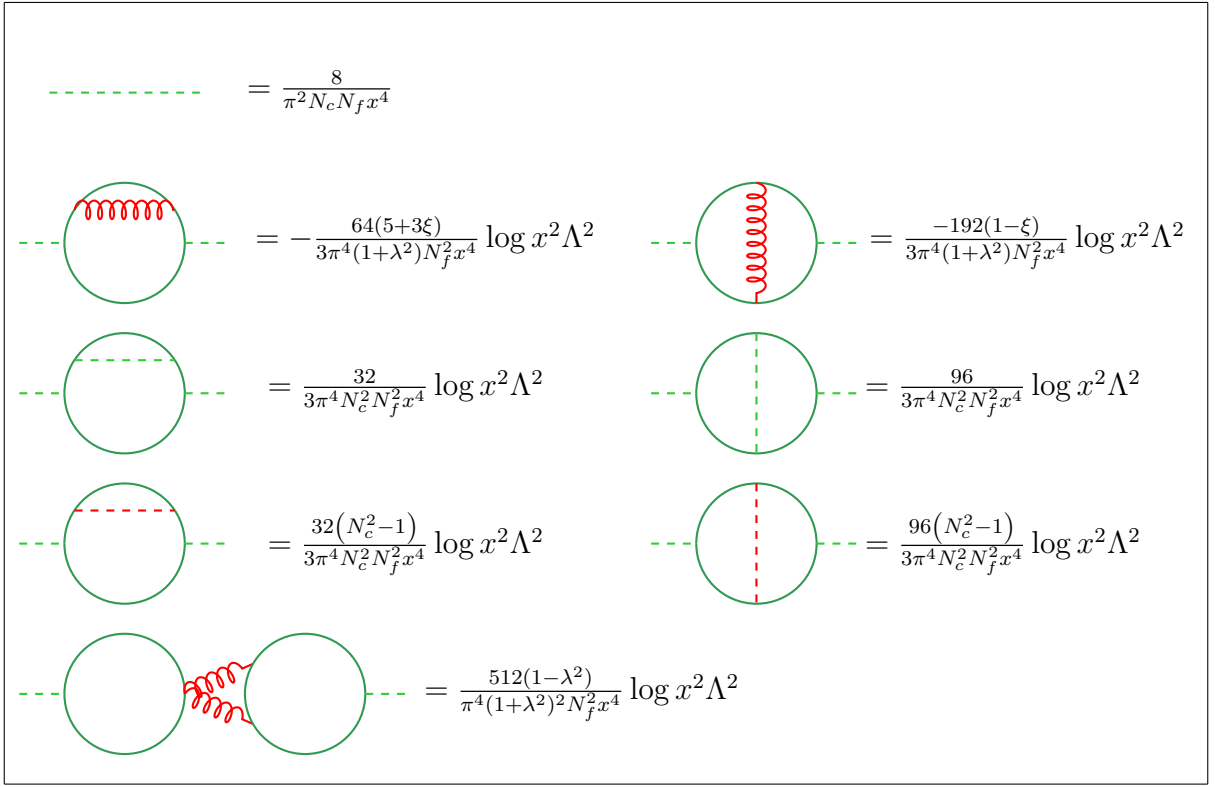
\begin{figure}[h!]
	\centering
	\begin{tikzpicture}[scale=0.4]
	
\draw[black] (-4-5-5,3) to (21+5,3);
\draw[black] (-4-5-5,3) to (-4-5-5,-40+2+5+10);
\draw[black] (21+5,3) to (21+5,-40+2+5+10);
\draw[black] (-4-5-5,-40+2+5+10) to (21+5,-40+2+5+10);

\draw[dashed,mygreen,thick] (-3.5+1-5-5,0) to (1.5+1-5-5,0);
\node at (10-0.5-5-5-3,0) {$= \frac{8}{\pi^2 N_c N_f x^4}$};

\draw[greenZS,thick] ([shift=(180:2cm)]0-5-5,-5)   arc (180:0:2cm);
\draw[greenZS,thick] ([shift=(0:2cm)]0-5-5,-5)   arc (0:-180:2cm);
\draw[dashed,mygreen,thick] (-3.5-5-5,-5) to (-2-5-5,-5);
\draw[dashed,mygreen,thick] (2-5-5,-5) to (3.5-5-5,-5);
\draw[gluon,red] (-1.66-5-5,-3.9) to (1.66-5-5,-3.9);
\node at (10-5-5,-5) {$ =-\frac{64(5+3\xi)}{3\pi^4(1+\lambda^2)N_f^2x^4}\log x^2\Lambda^2$};

\draw[greenZS,thick] ([shift=(180:2cm)]10+2-1,-5)   arc (180:0:2cm);
\draw[greenZS,thick] ([shift=(0:2cm)]10+2-1,-5)   arc (0:-180:2cm);
\draw[gluon,red] (10+2-1,-0) ([shift=(90:2cm)]10+2-1,-5)  to   ([shift=(-90:2cm)]10+2-1,-5);
\draw[dashed,mygreen,thick] (-3.5+12-1,-5) to (-2+12-1,-5);
\draw[dashed,mygreen,thick] (2+12-1,-5) to (3.5+12-1,-5);
\node at (15+5,-5) {$=\frac{-192(1-\xi)}{3\pi^4(1+\lambda^2)N_f^2x^4}\log x^2\Lambda^2$};

	\draw[greenZS,thick] ([shift=(180:2cm)]0-5-5,-10)   arc (180:0:2cm);
\draw[greenZS,thick] ([shift=(0:2cm)]0-5-5,-10)   arc (0:-180:2cm);
\draw[dashed, thick,mygreen] (-1.66-5-5,-8.9) to (1.66-5-5,-8.9);
\draw[dashed,mygreen,thick] (-3.5-5-5,-5-5) to (-2-5-5,-5-5);
\draw[dashed,mygreen,thick] (2-5-5,-5-5) to (3.5-5-5,-5-5);
\node at (8.05-5-0.7-3,-10) {$ =\frac{32}{3\pi^4N_c^2N_f^2 x^4}\log x^2\Lambda^2$};

\draw[greenZS,thick] ([shift=(180:2cm)]0+12-1,-10)   arc (180:0:2cm);
\draw[greenZS,thick] ([shift=(0:2cm)]0+12-1,-10)   arc (0:-180:2cm);
\draw[dashed,thick,mygreen] (0+12-1,-5) ([shift=(90:2cm)]0+12-1,-10)  to   ([shift=(-90:2cm)]0+12-1,-10);
\draw[dashed,mygreen,thick] (-3.5+12-1,-5-5) to (-2+12-1,-5-5);
\draw[dashed,mygreen,thick] (2+12-1,-5-5) to (3.5+12-1,-5-5);
\node at (7.9+12-0.7,-10) {$=\frac{96}{3\pi^4N_c^2N_f^2x^4}\log x^2\Lambda^2$};

\draw[greenZS,thick] ([shift=(180:2cm)]0-5-5,-15)   arc (180:0:2cm);
\draw[greenZS,thick] ([shift=(0:2cm)]0-5-5,-15)   arc (0:-180:2cm);
\draw[dashed,thick,red] (-1.66-5-5,-8.9-5) to (1.66-5-5,-8.9-5);
\draw[dashed,mygreen,thick] (-3.5-5-5,-5-5-5) to (-2-5-5,-5-5-5);
\draw[dashed,mygreen,thick] (2-5-5,-5-5-5) to (3.5-5-5,-5-5-5);
\node at (8.05-5-0.7-3,-15) {$ =\frac{32\left(N_c^2-1\right)}{3\pi^4N_c^2N_f^2x^4}\log x^2\Lambda^2$};

\draw[greenZS,thick] ([shift=(180:2cm)]0+12-1,-15)   arc (180:0:2cm);
\draw[greenZS,thick] ([shift=(0:2cm)]0+12-1,-15)   arc (0:-180:2cm);
\draw[dashed,thick,red] (0+12-1,-5) ([shift=(90:2cm)]0+12-1,-15)  to   ([shift=(-90:2cm)]0+12-1,-15);
\draw[dashed,mygreen,thick] (-3.5+12-1,-5-5-5) to (-2+12-1,-5-5-5);
\draw[dashed,mygreen,thick] (2+12-1,-5-5-5) to (3.5+12-1,-5-5-5);
\node at (7.9+12-0.7,-15) {$=\frac{96\left(N_c^2-1\right)}{3\pi^4N_c^2N_f^2x^4}\log x^2\Lambda^2$};

\draw[thick,greenZS]  ([shift=(0:2cm)]0-10,-20)  arc (0:180:2cm);
\draw[thick,greenZS]  ([shift=(-180:2cm)]0-10,-20) arc (-180:0:2cm);
\draw[gluon, red] (0-10,-20) ([shift=(0:2cm)]0-10,-20)  to   ([shift=(-210:2cm)]6-10,-20);
\draw[gluon, red] (0-10,-20) ([shift=(0:2cm)]0-10,-20)  to   ([shift=(-150:2cm)]6-10,-20);
\draw[thick,greenZS ] ([shift=(180:2cm)]6-10,-20) arc (180:0:2cm);
\draw[thick,greenZS] ([shift=(0:2cm)]6-10,-20) arc (0:-180:2cm);
\draw[dashed,mygreen,thick] (-3.5-10,-20) to (-2-10,-20);
\draw[dashed,mygreen,thick] (8-10,-20) to (9.5-10,-20);  
\node at (14.2-10+1,-20) {$= \frac{512(1-\lambda^2)}{\pi^4(1+\lambda^2)^2N_f^2x^4}\log x^2\Lambda^2$};

	\end{tikzpicture}
	\caption{ (bQCD) Results for individual Feynman graphs appearing in the 2-point correlation function of the singlet Hubbard-Stratonovich field. } \label{bQCDHS}
\end{figure}

The result of the logarithmic divergences for the contributing diagrams to $\langle \phi^2_{\text{adj}}(x) \phi^2_{\text{adj}}(0)\rangle$ have been sumarized in Figure \ref{bQCDadjfeynman}, while the corresponding ones for $\langle \sigma_s(x)\sigma_s(0) \rangle$ are in Figure \ref{bQCDHS}. The final result for the scaling dimension of the operators is
\begin{align}
   & \Delta[\phi^2_{\text{adj}}] = 1 + \frac{16 N_c (\lambda^2-3)}{3\pi^2(1+\lambda^2) N_f}\,, \\
  &  \Delta[\sigma_s] = 2 - \frac{16 N_c (\lambda^4-14\lambda^2+9)}{3\pi^2 (1+\lambda^2)^2 N_f}\,.
\end{align}

\begin{figure}[h!]
    \centering
    \includegraphics[width=0.6\textwidth]{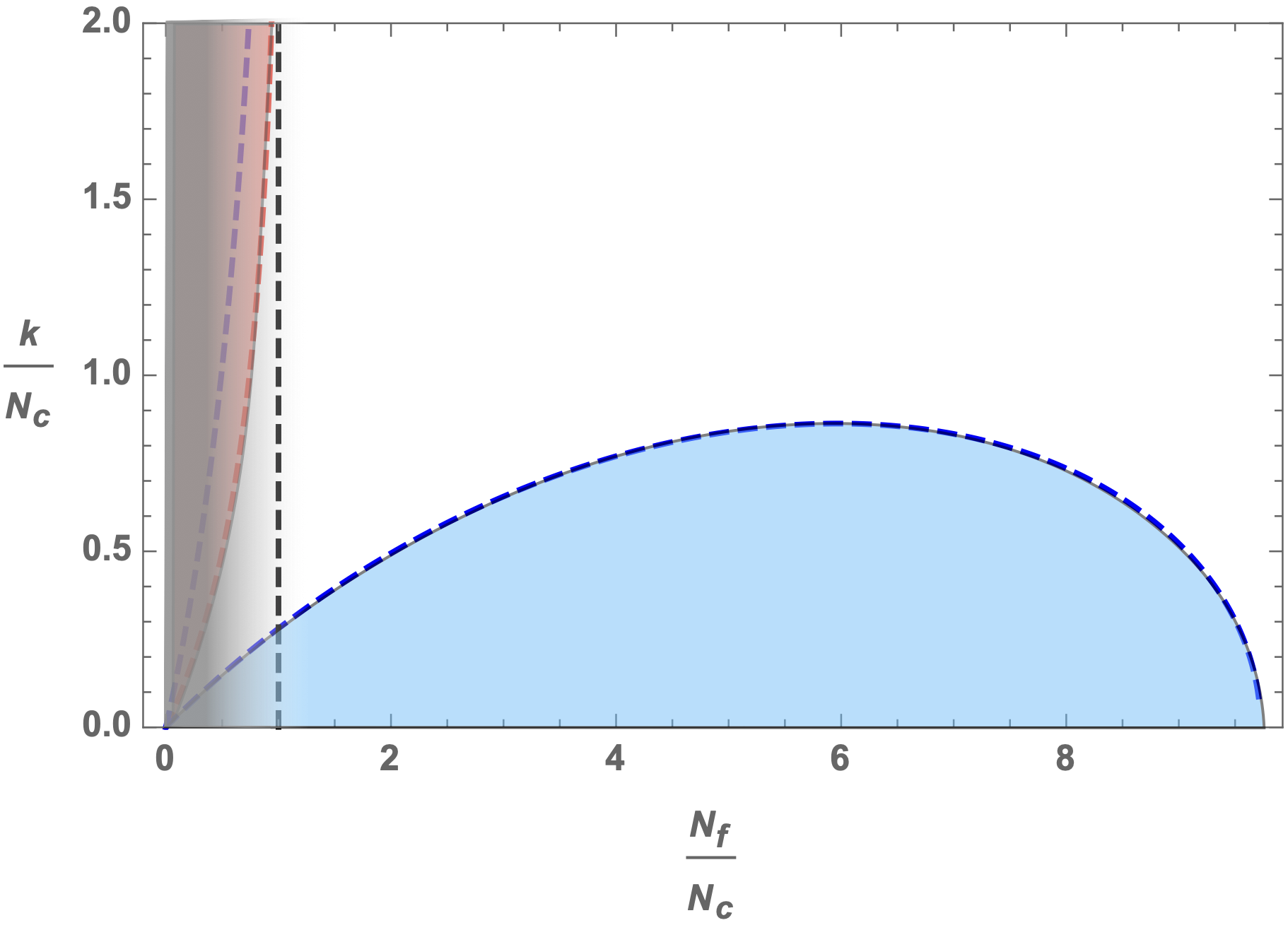}
    \caption{Locus in parameter space where the operators $(\phi^2_{\text{adj}})^2$ (red dashed line) and $\sigma_s^2$ (blue dashed line) hit marginality. The line $N_f/N_c = 1$ is drawn in black. The shaded region represents the region where our results are not applicable.}
    \label{fig:bosonic_merging_locus}
\end{figure}

With these results, we can proceed to analyze the resulting parameter space, similarly to the fermionic QCD case of the previous section. It is worth first pausing to discuss the expectations from bosonization dualities. To that matter, let us consider $\mathrm{SU}(N_c')_{k'}$ (fermionic) QCD$_3$ with $N_f'$ flavors. For $k'\geq \frac{N_f'}{2}$, where at $m=0$ the global $\mathrm{U}(N_f')$ symmetry is expected to be preserved, we can use duality, and describe the system by $\mathrm{U}(\frac{N_f'}{2}+{k'})_{-N_c'}+N_f'\,\phi$. Defining
\begin{equation}
    N_c=\frac{N_f'}{2}+k'\,,\qquad k=N_c'\,,\qquad N_f=N_f'\,,
\end{equation}
the dual theory is $\mathrm{U}(N_c)_{-k}+N_f\,\phi$, which, for $N_c\gg 1$ and $N_f\gg 1$, can be approximated by our computation for bQCD$_3$. Note that this requires $\frac{k}{N_c}$ to be small so that $N_f> N_c$ (at least parametrically). In terms of the unprimed variables, the regime of validity for the duality is $ \frac{N_f}{N_c}\leq 1$. Thus, making use of the duality in this regime, we expect bQCD$_3$ not to break the global $\mathrm{U}(N_f)$ symmetry. On the other hand, for $ \frac{N_f}{N_c}\geq 1$, we don't have any \textit{a priori} expectation. With this on mind, in Figure  \ref{fig:bosonic_merging_locus}, we depict the locus in parameter space where the operators hit marginality, corresponding to $\Delta[\phi^2_{\text{adj}}],\,\Delta[\phi^2_{\text{sing}}]=\frac{3}{2}$. In the bottom right of the diagram, and focusing to the right of the vertical line corresponding to $\frac{N_f}{N_c}=1$, there is a region (depicted in blue) where $\sigma_s$ hits marginality. At $k=0$, we find that the critical value of $N_f$ is
\begin{align}
    \frac{N_f}{N_c}\Big|_{k=0} \sim 9.7\,.
\end{align}
The region on the interior of this curve corresponds to a phase where $\sigma_s$ becomes marginal. Just as in the (fermionic) QCD$_3$ case, it can either be that there is fixed point crossing (between bQCD$_3$ and bQCD$_3^*$, corresponding to bQCD$_3$ deformed by $\sigma_s^2$) or annihilation. In the first scenario the blue region would correspond to a CFT. In the second scenario it is natural to expect that a VEV for $\sigma_s$ is generated, rendering the $\phi$ massive. As a consequence, in the IR we would be left with the TQFT $\mathrm{SU}(N_c)_k$. However, shifting the bare mass of the scalars we could offset this generated mass and remain at a conformal phase. In any case, the $\mathrm{U}(N_f)$ global symmetry would be preserved. Thus, the conclusion is that to the right of the dashed line corresponding to $\frac{N_f}{N_c}=1$ the $\mathrm{U}(N_f)$ global symmetry remains unbroken, and, possibly after tuning the bare mass of the $\phi$, the system is in a conformal phase.

On the other hand, in the region to the left of the vertical line corresponding to $\frac{N_f}{N_c}=1$ we have the expectation that the $\mathrm{U}(N_f)$ global symmetry should remain unbroken. Recall that this expectation is only valid for $k\ll N_c$, precisely inside the blue region extending to the left of the vertical line in Figure \ref{fig:bosonic_merging_locus}. As $k$ is increased, our computation seems to suggest that eventually the $\mathrm{U}(N_f)$ symmetry is broken (because the red region is hit first when decreasing $N_f$). Note however that the blue dashed line and the red dashed line are very close, so higher orders may affect this observation. In any case, this region is beyond the regime of validity of our computation. It would be interesting to study this part of the phase diagram from a large $k$, finite $N_f$ approximation, which we leave for future work.

\section{Conclusions and outlook}\label{sec:conclusions}

In this work we have studied massless QCD$_3$ with $N_f$ flavors. By working at large $N_f$, where the theory is conformal, one can develop a systematic appriximation in $\frac{1}{N_f}$. This allows to search for instabilities in the form of dangerously irrelevant operators. The natural candidates for those are quartics in the fermions. By working in large $N_c$ (albeit much smaller than $N_f$ in order not to enter the Veneziano regime) this is greatly simplified: there are only two candidate quartics. Since these are respectively the square of a flavor singlet and a flavor adjoint gauge-invariant, large $N_c$ factorization allows us to solely look to the dimensions of these bilinears which moreover, being in a different flavor representation, cannot mix. This simplifies the computation and allows to search for the locus in parameter space where these operators become marginal. It turns out that at $\frac{2k}{N_f}\sim 0.78$ the dangerously irrelevant operator changes from being the square of a flavor adjoint (for $\frac{2k}{N_f}\lsim 0.78$) to being the square of a flavor singlet (for $\frac{2k}{N_f}\gsim 0.78$); happening for $\frac{N_f}{N_c}\sim 2.3$ which is, at least parametrically, within the regime of validity of our approximation. As argued in the text, the presence of the dangerously irrelevant operator signals the nearby QCD$_3^*$ fixed point. It is natural then to assume fixed point annihilation, which results in conformality loss and, naturally, in the spontaneous symmetry breaking of the flavor symmetry for $\frac{2k}{N_f}\lsim 0.78$ in qualitative agreement with the Vafa-Witten theorem as well as with  \cite{Komargodski:2017keh}. Our analysis predicts a region --depicted in blue in Figure \ref{fig:ph_diagV4_Annotated}-- of broken conformal symmetry while unbroken flavor symmetry for $\frac{2k}{N_f}\gsim 0.78$. 
All these observations result in our conjectured phase diagram for massless QCD$_3$ in \ref{fig:ph_diagV4_Annotated}.

Concentrating on $k=0$, for fixed $N_c$ we find a critical $N_f$ at which the flavor-adjoint squared becomes marginal, which we interpreted as signaling the annihilation of QCD and QCD$_3^*$ resulting in the spontaneous symmetry breaking of $\mathrm{U}(N_f)$ into $\mathrm{U}\left(\frac{N_f}{2}\right)\times \mathrm{U}\left(\frac{N_f}{2}\right)$ as prescribed by the Vafa-Witten theorem. This motivated to study other quartic operators which, although not singlets under $\mathrm{U}(N_f)$, contain a singlet when branched under $\mathrm{U}\left(\frac{N_f}{2}\right)\times \mathrm{U}\left(\frac{N_f}{2}\right)$. We have seen that also these go towards marginality at the precise same critical value of $N_f$.

Even though our computation assumes $N_f\gg N_c\gg 1$, we have extrapolated our results to $\frac{N_f}{N_c}\sim 4$, which is only parametrically inside the regime of validity. However, we find a consistent emerging picture. Moreover, in the $k=0$ region where higher $\frac{1}{N_f}$ corrections can be included, we find that these shift the critical $N_f$ for the fixed point annihilation towards larger values, that is, deeper in to the regime of validity of our approximation, thus giving confidence on the emerging picture.

We have conjectured that the line separating the red/blue regions from the white region in  \ref{fig:ph_diagV4_Annotated} corresponds to fixed point annihilation. However, it would be very interesting to further study this point. Note also that for small values of $\frac{N_f}{N_c}$ we would enter the regime of validity of \cite{Armoni:2019lgb}. It would be very interesting to study the transition between these regimes. Along these lines, it would be very interesting to clarify the role of the quartic non-singlet operators which, seemingly, hit marginality at the same critical value for $N_f$.

The particular case of $\mathrm{SU}(N_c)$ fermionic QCD$_3$ with $k=1-\frac{N_f}{2}$ can be regarded as the domain wall separating the two vacua of QCD$_4$ with $\theta=\pi$. From this point of view, it seems natural to conjecture a connection between the conformality/non-conformality of QCD$_4$ with that of the domain wall theory. Indeed, the critical value for $N_f$ that we find is very close to the lower bound of the conformal window for QCD$_4$. It would be very interesting to explore this hypothetical connection.

We have also studied the bosonic counterpart bQCD$_3$ using the same strategy. In this case, in the regime of validity of our approximation, to the right of the dashed vertical line in Figure \ref{fig:bosonic_merging_locus}, we find a region of conformal symmetry breaking but no flavor symmetry breaking below some curve conjecturally triggered by fixed point annihilation.

\section*{Acknowledgements}

It is a pleasure to thank Hrachya Khachatryan for discussions, as well as Marco Serone for useful conversations and careful reading of the draft. GAT would also like to thank the hospitality of SISSA through their ``Visiting PhD training program" at the beginning of this project. The work of GAT and DRG is partly supported by Spanish national grant MCIU-22-PID2021-123021NB-I00 as well as the Principado de Asturias grant SV-PA-21-AYUD/2021/52177. GAT is also supported by the Spanish government scholarship MCIU-19-FPU18/02221. SB is partially supported by the INFN Research Project GAST.

\begin{appendix}

\section{Scaling dimension of bilinear operators in fermionic CS-QCD at large $N_f$}\label{app:computation_singlets}

In this appendix we compile the relevant details of the computation of the anomalous dimensions of the two bilinear operators $\psi^2_{\text{adj}}$ and $\psi^2_{\text{sing}}$ from the large $N_f$ expansion. In order to do that, we employ the same strategy as \cite{Chester:2016ref}, with the Feynman rules of Figure \ref{fig:fQCDfeynmanrules}. The first step is to write down all the diagrams that contribute up to order $1/N_f$. Having performed the resummation of Figure \ref{fig:effective_gluon}, this becomes very easy: the effective gluon propagator contributes as $1/N_f$, while fermion loops contribute as $N_f$. 

At tree level there is only one diagram, which equals
\begin{align}\label{eq:feynman_tree_level_fqcd}
    \begin{tikzpicture}[baseline={([yshift=-.5ex]current bounding box.center)},scale=0.5]
    \draw[thick, blue] (0,0) circle (2cm);
	\draw[thick,fill] (-2,0) circle (4pt);
	\draw[thick,fill] (2,0) circle (4pt);
	\node at (-2.6,0) {$x$};
	\node at (2.6,0) {$0$};
    \end{tikzpicture} \, =\, - \frac{N_fN_c}{8\pi^2x^4}\,.
\end{align}
Note that we have chosen the normalization of $\psi^2_{\text{sing}}$ in \eqref{eq:psi2sing} in such a way that \eqref{eq:feynman_tree_level_fqcd} holds for both bilinear operators.

At order $1/N_f$ we have two diagrams given by adding a gluon line to the one in \eqref{eq:feynman_tree_level_fqcd}. These are the second and third diagrams in Figure \ref{fQCDbilinears}. Once again, thanks to the choice of normalization, they contribute equally to $\psi^2_{\text{sing}}$ and $\psi^2_{\text{adj}}$. The last diagram that contributes at the same order consists of two fermion loops joined by two gluons; it is the fourth diagram in Figure \ref{fQCDbilinears}. Note that this diagram only contributes to $\psi^2_{\text{sing}}$, and not to $\psi^2_{\text{adj}}$, due to the contraction of the flavour indices in the fermion loops.

As described in the main text, the full computation of these Feynman diagrams can be quite involved. Thankfully, we are only interested on extracting the anomalous dimension of the operators, and therefore we only need to compute the coefficient of the logarithmic divergence.

Let's begin by considering the following diagram in position space,
\begin{align}
        \begin{tikzpicture}[baseline={([yshift=-.5ex]current bounding box.center)},scale=0.4]
	\draw[blue,thick] ([shift=(180:2cm)]0,-10)   arc (180:0:2cm);
	\draw[blue,thick] ([shift=(0:2cm)]0,-10)   arc (0:-180:2cm);
	\draw[gluon, red] (0,-10) ([shift=(90:2cm)]0,-10)  to   ([shift=(-90:2cm)]0,-10);
	\draw[thick,fill] (-2,-10) circle (4pt);
	\draw[thick,fill] (2,-10) circle (4pt);
	\node at (-2.6,-10) {$x$};
	\node at (2.6,-10) {$0$};
	\node at (0,-7.5) {$z$};
	\node at (0,-12.5) {$w$};
        \end{tikzpicture}
        = - N_f (N_c^2-1) \int d^3z\, d^3w\, D_{\mu\nu}(z,w)\,\text{Tr}\left[G(x,z)\gamma^\mu G(z,0) G(0,w)\gamma^\nu G(w,x)  \right]\,,
\end{align}
where $D_{\mu\nu}(z,w)$ denotes the effective gluon propagator in position space, whose explicit expression we won't need, except to note that it's dependence is $\propto |z-w|^{-2}$. By counting powers of $z$ and $w$ we can see that the logarithmic divergence comes from taking the limit where $z$ and $w$ are both close to $x$ or to $0$ at the same time. In fact, both divergences turn out to give the same result, so we can take $z,w\to0$ and multiply by two, 
\begin{align}
        \begin{tikzpicture}[baseline={([yshift=-.5ex]current bounding box.center)},scale=0.4]
	\draw[blue,thick] ([shift=(180:2cm)]0,-10)   arc (180:0:2cm);
	\draw[blue,thick] ([shift=(0:2cm)]0,-10)   arc (0:-180:2cm);
	\draw[gluon, red] (0,-10) ([shift=(90:2cm)]0,-10)  to   ([shift=(-90:2cm)]0,-10);
	\draw[thick,fill] (-2,-10) circle (4pt);
	\draw[thick,fill] (2,-10) circle (4pt);
	\node at (-2.6,-10) {$x$};
	\node at (2.6,-10) {$0$};
	\node at (0,-7.5) {$z$};
	\node at (0,-12.5) {$w$};
        \end{tikzpicture}
      &\simeq -2 N_f(N_c^2-1) \int d^3z\, d^3w\, D_{\mu\nu}(z,w) \, \text{Tr}\left[ G(0,x) G(x,0) \gamma^\mu G(z,0) G(0,w)\gamma^\nu \right]\nonumber \\
      & = \frac{2 N_f (N_c^2-1)}{8\pi^2 x^4} \underbrace{\int d^3z\, d^3w\, D_{\mu\nu}(z,w)\text{Tr}\left[ \gamma^\mu G(z,0) G(0,w) \gamma^\nu \right]}_{\mathcal{I}}\,.
\end{align}
In order to compute $\mathcal{I}$, we Fourier transform to go to momentum space,
\begin{align}
    \mathcal{I} &= \int \frac{d^3p}{(2\pi)^3}\, D_{\mu\nu}(p)\, \text{Tr}\left[ \gamma^\mu G(p) G(p) \gamma^\nu \right]\\
    &=-\int \frac{d^3p}{(2\pi)^3}\, D_{\mu\nu}(p)\frac{p_\alpha p_\beta}{p^4}\,\text{Tr}\left[\gamma^\mu\gamma^\alpha\gamma^\beta\gamma^\nu\right]\,.
\end{align}
After some algebra with the gamma matrices, this results in
\begin{align}\label{eq:D1}
    \mathcal{I}=- 2\delta^{\mu\nu} \int \frac{d^3p}{(2\pi)^3}\, \frac{D_{\mu\nu}(p)}{p^2} = -\frac{8(3-\xi)}{\pi^2(1+\lambda^2)N_f} \log x^2\Lambda^2\,,
\end{align}
where $\Lambda$ is the high-energy cutoff in the integral. All in all, we have
\begin{align}
        \begin{tikzpicture}[baseline={([yshift=-.5ex]current bounding box.center)},scale=0.4]
	\draw[blue,thick] ([shift=(180:2cm)]0,-10)   arc (180:0:2cm);
	\draw[blue,thick] ([shift=(0:2cm)]0,-10)   arc (0:-180:2cm);
	\draw[gluon, red] (0,-10) ([shift=(90:2cm)]0,-10)  to   ([shift=(-90:2cm)]0,-10);
	\draw[thick,fill] (-2,-10) circle (4pt);
	\draw[thick,fill] (2,-10) circle (4pt);
	\node at (-2.6,-10) {$x$};
	\node at (2.6,-10) {$0$};
        \end{tikzpicture}
       = - \frac{N_c^2-1}{\pi^4 (1+\lambda^2)x^4}(3-\xi)\log x^2\Lambda^2\,.
\end{align}

The second diagram with one gluon line is easier to compute, as the self-energy of the fermion coming from the effective gluon has already been computed in \cite{Rantner:2002zz},
\begin{align}
        \begin{tikzpicture}[baseline={([yshift=-.5ex]current bounding box.center)},scale=0.4]
        \draw[gluon,red] ([shift=(180:2cm)]0,-10)   arc (180:0:2cm);
        \draw[blue,thick] (-3.5,-10)--(3.5,-10);
        \end{tikzpicture} 
        = -\frac{4(N_c^2-1)}{\pi^2 N_c (1+\lambda^2) N_f }\left(\frac{1}{3}-\xi\right) G(x,0)\log x^2\Lambda^2\,.
\end{align}
This results inmediately in
\begin{align}\label{eq:D2}
    \begin{tikzpicture}[baseline={([yshift=-.5ex]current bounding box.center)},scale=0.4]
    	\draw[blue,thick] ([shift=(180:2cm)]0,-5)   arc (180:0:2cm);
	\draw[blue,thick] ([shift=(0:2cm)]0,-5)   arc (0:-180:2cm);
	\draw[gluon,red] (-1.66,-3.9) to (1.66,-3.9);
	\draw[thick,fill] (-2,-5) circle (4pt);
	\draw[thick,fill] (2,-5) circle (4pt);
    \end{tikzpicture}
    = \frac{(N_c^2-1)}{\pi^4(1+\lambda^2)x^4}\left(\frac{1}{3}-\xi\right)\log x^2\Lambda^2\,.
\end{align}

The last diagram, which only contributes to the two-point function of the singlet quadratic operator, is
\begin{align}\label{eq:diagram_fqcd_long}
    \begin{tikzpicture}[baseline={([yshift=-.5ex]current bounding box.center)},scale=0.4]
    	\draw[thick,blue]  ([shift=(0:2cm)]0,-15)  arc (0:180:2cm);
	\draw[thick,blue]  ([shift=(-180:2cm)]0,-15) arc (-180:0:2cm); 
	\draw[gluon, red] ([shift=(40-14:2cm)]0,-15)  to   ([shift=(140+14:2cm)]6,-15);
	\draw[gluon, red] ([shift=(-40+14:2cm)]0,-15)  to   ([shift=(-140-14:2cm)]6,-15);
	\draw[thick,blue ] ([shift=(180:2cm)]6,-15) arc (180:0:2cm);
	\draw[thick,blue] ([shift=(0:2cm)]6,-15) arc (0:-180:2cm);
	\draw[thick,fill] (-2,-15) circle (4pt);
	\draw[thick,fill] (8,-15) circle (4pt);
	\node at (-2.6,-15) {$x$};
	\node at (8.6,-15) {$0$};
	\node at (0.8,-14) {$y_1$};
	\node at (0.8,-16) {$y_2$};
	\node at (5.25,-14) {$y_3$};
	\node at (5.25,-16) {$y_4$};
    \end{tikzpicture}
    =& N_f^2 (N_c^2-1) \int d^3y_1 d^3y_2d^3y_3d^3y_4\, \text{Tr}\left[G(x,y_1)\gamma^\mu G(y_1,y_2)\gamma^\alpha G(y_2,x) \right]\nonumber\\
    &\times D_{\mu\nu}(y_1,y_3) D_{\alpha\beta}(y_2,y_4) \text{Tr}\left[G(0,y_4)\gamma^\beta G(y_4,y_3)\gamma^\nu G(y_3,0) \right]\,.
\end{align}
Similarly to the first diagram, the logarithmic divergence can be extracted from taking the limit $y_1,y_2\to 0$, or $y_3,y_4\to x$, which results in
\begin{align}
    \eqref{eq:diagram_fqcd_long} = -\frac{2N_f (N_c^2-1)}{8\pi^2 x^4} \int d^3y_1 d^3y_2d^3y_3d^3y_4\, &\text{Tr}\left[\gamma^\mu G(y_1,y_2)\gamma^\alpha\right] D_{\mu\nu}(y_1,y_3) D_{\alpha\beta}(y_2,y_4)\nonumber\\
   & \times\text{Tr}\left[G(0,y_4)\gamma^\beta G(y_4,y_3)\gamma^\nu G(y_3,0) \right]\,. 
\end{align}
After Fourier transforming, this becomes
\begin{align}
-\frac{2N_f^2(N_c^2-1)}{8\pi^2 x^4}\int\frac{d^3p}{(2\pi)^3}\frac{d^3k}{(2\pi)^3}\text{Tr}\left[\gamma^\mu G(p)\gamma^\alpha\right] D_{\mu\nu}(-p) D_{\alpha\beta}(p)\text{Tr}\left[ G(k)\gamma^\beta G(p-k)\gamma^\nu G(k) \right]\,.   
\end{align}
We begin by doing the integral in $k$, which only involves the trace coming from the second loop of fermions,
\begin{align}
    \mathcal{I}_k=\int \frac{d^3k}{(2\pi)^3}\frac{1}{k^2}\text{Tr}[\gamma^\beta G(p-k)\gamma^\nu] = -i\frac{p_\sigma \epsilon^{\beta\sigma\nu}}{8|p|}\,.
\end{align}
We can anticipate that the presence of the epsilon tensor, coming from the loop integral involving three fermion propagators, is what ultimately will result in a non-trivial contribution of the Chern-Simons term in the effective gluon propagator.

Next, we plug this result in \eqref{eq:diagram_fqcd_long}, and compute the second trace of the gamma matrices. This results in
\begin{align}
    \frac{N_f^2(N_c^2-1)}{16\pi^2 x^4} \int \frac{d^3p}{(2\pi)^3} \frac{p_\sigma p_\rho}{|p|^3} D_{\mu\nu}(-p) D_{\alpha\beta}(p)\epsilon^{\beta\sigma\nu}\epsilon^{\mu\rho\alpha}\,.
\end{align}
At this stage, it is necessary to use the explicit expression of the effective gluon propagator, and carry out an algebra exercise involving the epsilon matrices. In doing so, one sees that all the contributions involving the gauge fixing parameter $\xi$ cancel, while there is a non-vanishing term involving $\lambda$. The result is
\begin{align}\label{eq:result_fqcd_long}
    \begin{tikzpicture}[baseline={([yshift=-.5ex]current bounding box.center)},scale=0.4]
    	\draw[thick,blue]  ([shift=(0:2cm)]0,-15)  arc (0:180:2cm);
	\draw[thick,blue]  ([shift=(-180:2cm)]0,-15) arc (-180:0:2cm); 
	\draw[gluon, red] ([shift=(40-14:2cm)]0,-15)  to   ([shift=(140+14:2cm)]6,-15);
	\draw[gluon, red] ([shift=(-40+14:2cm)]0,-15)  to   ([shift=(-140-14:2cm)]6,-15);
	\draw[thick,blue ] ([shift=(180:2cm)]6,-15) arc (180:0:2cm);
	\draw[thick,blue] ([shift=(0:2cm)]6,-15) arc (0:-180:2cm);
	\draw[thick,fill] (-2,-15) circle (4pt);
	\draw[thick,fill] (8,-15) circle (4pt);
	\node at (-2.6,-15) {$x$};
	\node at (8.6,-15) {$0$};
    \end{tikzpicture}
    = \frac{4(N_c^2-1)(\lambda^2-1)}{\pi^4 (1+\lambda^2)^2 x^4}\log x^2\Lambda^2\,.
\end{align}

Note that the gauge-dependence in \eqref{eq:D1} and \eqref{eq:D2} exactly cancels. On the other hand, eq. \eqref{eq:result_fqcd_long} is gauge-indepedent, as it should given that it only contributes to the two-point function of the flavor singlet, and thus directly corresponds to the difference between two anomalous dimensions and is therefore a physical observable.

\section{Scaling dimension of flavor non-singlet quartic operators}\label{app:computation_non_singlets}

In this appendix we collect the details of the computation of the scaling dimension of the quartic flavor non-singlet operators, for fermionic QCD$3$ at vanishing CS level, introduced in the main text.

\subsubsection*{Correction to the fermion propagator}\label{app_self_energy}

Firstly, consider the $1/N_f$ correction to the fermion propagator, coming from the following diagram, and which equals,
\begin{align*}
\begin{tikzpicture}[scale=0.4]
\draw[scalar] (-3.5,0) to (3.5,0);
\draw[gluon, red] (0,0) (0:2) arc (0:180:2.2);
\node at (-7,0) {$\widetilde{G}_{ia}^{jb}(x,0)=$};
\node at (16.5,0) {$=G(x,0)\,\delta_i^j\left[-\frac{4}{\pi^2 N_f}\left(\frac{1}{3}-\xi\right)\left(T_A\right)_a^x\left(T_B\right)_x^b\delta^{AB}\right]\log\Lambda^2$};
\end{tikzpicture}
\end{align*}
We will use this intermediate result to evaluate several of the relevant diagrams immediately.

\subsubsection*{Double trace operators}

We begin with the calculation of the correlators between the double trace operators, one by one.

\subsubsection*{$\langle O_1^d(x)\bar{O}_1^d(0) \rangle=\langle O_2^d(x)\bar{O}_2^d(0) \rangle$}

First, we have 4 diagrams where the gluon line goes from one fermion to itself, which can be easily evaluated using the result in \ref{app_self_energy}
\begin{align*}
&\begin{tikzpicture}[scale=0.4]
\draw[scalar] (0,0) (0:2) arc (0:180:2 and 1);
\draw[scalar] (0,0) (-180:2) arc (-180:0:2 and 1);
\draw[scalar] (0,0) (0:2) arc (0:180:2);
\draw[scalar] (0,0) (-180:2) arc (-180:0:2);
\draw[thick,fill] (-2,0) circle (4pt);
\draw[thick,fill] (2,0) circle (4pt);
\draw[gluon,red] (-1.37,1.4) to (1.37,1.4); 
\node at (-3.4,0) {$4\,\times$};
\node at (14.65,0) {$=4\,\text{Tr}[\widetilde{G}^{c}_a(x,0)G_c^a(0,x)]\,\text{Tr}[G^b_d(x,0)G^d_b(0,x)]  \qquad\qquad$};
\end{tikzpicture} \\
&\hspace{2.75cm}=4\,\underbrace{\delta^d_b\delta^b_d}_{N_c}\underbrace{\delta^a_c(T^A)^x_a(T^B)^c_x\delta_{AB}}_{N_c^2-1}\left[-\frac{4}{\pi^2 N_f}\left(\frac{1}{3}-\xi\right)\right] \underbrace{\text{Tr}[G(x,0)G(0,x)]^2}_{\left(\frac{2}{(4\pi)^2x^4}\right)^2} \log \Lambda^2\\
&\hspace{2.75cm}=-\frac{N_c\left(N_c^2-1\right)}{4\pi^6N_fx^8}\left(\frac{1}{3}-\xi\right)\log\Lambda^2
\end{align*}
Next, we have 6 diagrams were the gluon line links different fermions,

\begin{align*}
&\begin{tikzpicture}[scale=0.4]
\draw[scalar] (0,0) (0:2) arc (0:180:2 and 1);
\draw[scalar] (0,0) (-180:2) arc (-180:0:2 and 1);
\draw[scalar] (0,0) (0:2) arc (0:180:2);
\draw[scalar] (0,0) (-180:2) arc (-180:0:2);
\draw[thick,fill] (-2,0) circle (4pt);
\draw[thick,fill] (2,0) circle (4pt);
\draw[gluon,red] (0,2) to (0,1); 
\node at (11.5,0) {$=-\mathop{\mathlarger{\int}} dz\,dw\,D_{\mu\nu}^{AB}(z,w)\qquad\qquad\qquad$};
\end{tikzpicture} \\
&\hspace{2.5cm}\times\text{Tr}[G_a^{x_1}(x,z)\gamma^\mu(T_A)_{x_1}^{x_2}G_{x_2}^c(z,0)G_c^a(0,x)]\,\text{Tr}[G_d^{y_1}(x,w)\gamma^\nu(T_B)_{y_1}^{y_2}G_{y_2}^b(w,0)G_b^d(0,x)]\\
&\hspace{2cm}=-\delta_a^{x_1}(T_A)_{x_1}^{x_2}\delta_{x_2}^c\delta_c^a\left(\cdots\right)\\
&\hspace{2cm} = 0
\end{align*}
because Tr$[T_A]=0$. For the same reason,
\begin{align*}
&\begin{tikzpicture}[scale=0.4]
\draw[scalar] (0,0) (0:2) arc (0:180:2 and 1);
\draw[scalar] (0,0) (-180:2) arc (-180:0:2 and 1);
\draw[scalar] (0,0) (0:2) arc (0:180:2);
\draw[scalar] (0,0) (-180:2) arc (-180:0:2);
\draw[thick,fill] (-2,0) circle (4pt);
\draw[thick,fill] (2,0) circle (4pt);
\draw[gluon,red] (0,2) to (0,-2); 
\node at (3,0) {$=$};
\end{tikzpicture}
\begin{tikzpicture}[scale=0.4]
\draw[scalar] (0,0) (0:2) arc (0:180:2 and 1);
\draw[scalar] (0,0) (-180:2) arc (-180:0:2 and 1);
\draw[scalar] (0,0) (0:2) arc (0:180:2);
\draw[scalar] (0,0) (-180:2) arc (-180:0:2);
\draw[thick,fill] (-2,0) circle (4pt);
\draw[thick,fill] (2,0) circle (4pt);
\draw[gluon,red] (0,1) to (0,-1); 
\node at (3,0) {$=$};
\end{tikzpicture} 
\begin{tikzpicture}[scale=0.4]
\draw[scalar] (0,0) (0:2) arc (0:180:2 and 1);
\draw[scalar] (0,0) (-180:2) arc (-180:0:2 and 1);
\draw[scalar] (0,0) (0:2) arc (0:180:2);
\draw[scalar] (0,0) (-180:2) arc (-180:0:2);
\draw[thick,fill] (-2,0) circle (4pt);
\draw[thick,fill] (2,0) circle (4pt);
\draw[gluon,red] (0,-2) to (0,-1); 
\node at (3.5,0) {$=0$};
\end{tikzpicture}
\end{align*}
The only remaining contribution comes from 

\begin{align*}
&\begin{tikzpicture}[scale=0.4]
\draw[scalar] (0,0) (0:2) arc (0:180:2 and 1);
\draw[scalar] (0,0) (-180:2) arc (-180:0:2 and 1);
\draw[scalar] (0,0) (0:2) arc (0:180:2);
\draw[scalar] (0,0) (-180:2) arc (-180:0:2);
\draw[thick,fill] (-2,0) circle (4pt);
\draw[thick,fill] (2,0) circle (4pt);
\draw[gluon,red] (0,2) to (0,-1); 
\node at (-3.5,0) {$2\,\times$};
\node at (15,0) {$=-2\mathop{\mathlarger{\int}} dz\,dw\,D_{\mu\nu}^{AB}(z,w)\,\text{Tr}[G_b^d(x,0)G^b_d(0,x)]\,\qquad\qquad$};
\end{tikzpicture}\\
&\hspace{3.5cm}\times\text{Tr}[G_a^{x_1}(x,0)\gamma^\mu(T_A)_{x_1}^{x_2}G_{x_2}^c(z,0)G_c^{y_1}(0,w)\gamma^\nu (T_B)_{y_1}^{y_2}G_{y_2}^a(w,x)]\\
&\hspace{2cm}\simeq -4 N_c\left(N_c^2-1\right)\int dz\,dw D_{\mu\nu}(z,w)\text{Tr}[G(x,0)G(0,x)]\,\text{Tr}[G(0,x)G(x,0)\gamma^\mu G(z,0)G(0,w)\gamma^\nu]\\
&\hspace{3cm}=-\frac{8N_c\left(N_c^2-1\right)}{(4\pi)^4x^8}\int dz\,dw\,D_{\mu\nu}(z,w)\text{Tr}\left[\gamma^\mu G(z,0)G(0,w)\gamma^\nu \right]\\
&\hspace{3cm}=\frac{N_c\left(N_c^2-1\right)}{4\pi^6N_fx^8}(3-\xi)\log\Lambda^2
\end{align*}
where in the second line we used that we are only interested in the divergent part, which comes from $z$ and $w$ at the same time close to $x$ or to $0$.

Summing both contributions,
\begin{align}
\boxed{\langle O_1^d(x)\bar{O}_1^d(0) \rangle=\langle O_2^d(x)\bar{O}_2^d(0) \rangle = \frac{2N_c\left(N_c^2-1\right)}{3\pi^6N_fx^8}\log\Lambda^2}
\end{align}

\subsubsection*{$\langle O_1^d(x)\bar{O}_2^d(0) \rangle$}

As before, first we have the 4 diagrams
\begin{align*}
&\begin{tikzpicture}[scale=0.4]
\draw[scalar] (0,0) (0:2) arc (0:180:2 and 1);
\draw[scalar] (0,0) (-180:2) arc (-180:0:2 and 1);
\draw[scalar] (0,0) (0:2) arc (0:180:2);
\draw[scalar] (0,0) (-180:2) arc (-180:0:2);
\draw[thick,fill] (-2,0) circle (4pt);
\draw[thick,fill] (2,0) circle (4pt);
\draw[gluon,red] (-1.37,1.4) to (1.37,1.4); 
\node at (-3.4,0) {$4\,\times$};
\node at (12,0) {$=-4\,\text{Tr}[\widetilde{G}^{c}_a(x,0)G_c^b(0,x)G^d_b(x,0)G^a_d(0,x)]  $};
\end{tikzpicture} \\
&\hspace{2.75cm}=\frac{N_c^2-1}{8\pi^6N_fx^8}\left(\frac{1}{3}-\xi\right)\log\Lambda^2
\end{align*}
where the difference in the contraction of Lorentz and colour indices comes from the Wick contraction of the fermions, which is essentially the same as in the tree level case.

In this operator, there is only one overall trace over the colour indices. Therefore, there will be no diagrams that cancel because of Tr$[T_A]=0$. There will be two different contributions, depending on whether the gluon links two fermions going in the same direction, or in different directions. The first one is
\begin{align*}
&\begin{tikzpicture}[scale=0.4]
\draw[scalar] (0,0) (0:2) arc (0:180:2 and 1);
\draw[scalar] (0,0) (-180:2) arc (-180:0:2 and 1);
\draw[scalar] (0,0) (0:2) arc (0:180:2);
\draw[scalar] (0,0) (-180:2) arc (-180:0:2);
\draw[thick,fill] (-2,0) circle (4pt);
\draw[thick,fill] (2,0) circle (4pt);
\draw[gluon,red] (0,2) to (0,1); 
\node at (-3,0) {$2\,\times\,$};
\node at (10,0) {$=2\mathop{\mathlarger{\int}} dz\,dw\,D_{\mu\nu}^{AB}(z,w)\qquad\qquad$};
\end{tikzpicture}\\
&\hspace{3.15cm}\times \text{Tr}\left[G_a^{x_1}(x,z)\gamma^\mu(T_A)_{x_1}^{x_2}G_{x_1}^c(z,0)G_{c}^b(0,x)G_{b}^{y_1}(x,w)\gamma^\nu(T_B)_{y_1}^{y_2}G_{y_2}^d(w,0)G_{d}^a(0,x)\right]\\
&\hspace{2.7cm}\simeq\frac{4\left(N_c^2-1\right)}{(4\pi)^4x^8}\int dz\,dw\, D_{\mu\nu}(z,w)\text{Tr}\left[\gamma^\mu G(z,0)\gamma^\nu G(w,0)\right]\\
&\hspace{2.7cm}=-\frac{N_c^2-1}{8\pi^6N_fx^8}(1+\xi)\log\Lambda^2
\end{align*}
where again we have used that we are interested only in the divergent part of the integral. The last contribution is
\begin{align*}
&\begin{tikzpicture}[scale=0.4]
\draw[scalar] (0,0) (0:2) arc (0:180:2 and 1);
\draw[scalar] (0,0) (-180:2) arc (-180:0:2 and 1);
\draw[scalar] (0,0) (0:2) arc (0:180:2);
\draw[scalar] (0,0) (-180:2) arc (-180:0:2);
\draw[thick,fill] (-2,0) circle (4pt);
\draw[thick,fill] (2,0) circle (4pt);
\draw[gluon,red] (0,2) to (0,-1); 
\node at (-3,0) {$4\,\times\,$};
\node at (10,0) {$=4\mathop{\mathlarger{\int}} dz\,dw\,D_{\mu\nu}^{AB}(z,w)\qquad\qquad$};
\end{tikzpicture}\\
&\hspace{3.1cm}\times \text{Tr}\left[G_a^{x_1}(x,z)\gamma^\mu(T_A)_{x_1}^{x_2}G_{x_2}^c(z,0)G_c^b(0,x)G_b^d(x,0)G_d^{y_1}(0,w)\gamma^\nu(T_B)_{y_1}^{y_2}G_{y_2}^a(w,x)\right]\\
&\hspace{2.7cm}\simeq\frac{8\left(N_c^2-1\right)}{(4\pi)^4x^8}\int dz\,dw\,D_{\mu\nu}(z,w)\text{Tr}[\gamma^\mu G(z,0)G(0,w)\gamma^\nu]\\
&\hspace{2.7cm}=-\frac{N_c^2-1}{4\pi^6N_fx^8}(3-\xi)\log\Lambda^2
\end{align*}
In total,
\begin{align}
\boxed{\langle O_1^d(x)\bar{O}_2^d(0)\rangle=-\frac{5\left(N_c^2-1\right)}{6\pi^6N_fx^8}\log\Lambda^2}
\end{align}

\subsubsection*{Single trace operators}

The calculation for the correlators of single trace operators proceeds in a similar way to the double trace ones, although different integrals appear in the diagrams.

\subsubsection*{$\langle O_1^s(x)\bar{O}_1^s(0) \rangle=\langle O_2^s(x)\bar{O}_2^s(0) \rangle$}

As always, we start with
\begin{align*}
&\begin{tikzpicture}[scale=0.4]
\draw[scalar] (0,0) (0:2) arc (0:180:2 and 1);
\draw[scalar] (0,0) (-180:2) arc (-180:0:2 and 1);
\draw[scalar] (0,0) (0:2) arc (0:180:2);
\draw[scalar] (0,0) (-180:2) arc (-180:0:2);
\draw[thick,fill] (-2,0) circle (4pt);
\draw[thick,fill] (2,0) circle (4pt);
\draw[gluon,red] (-1.37,1.4) to (1.37,1.4); 
\node at (-3.4,0) {$4\,\times$};
\node at (13,0) {$=-4\,\text{Tr}[\widetilde{G}^{d}_a(x,0)G_c^b(0,x)]\,\text{Tr}[G^c_b(x,0)G^a_d(0,x)]  $};
\end{tikzpicture} \\
&\hspace{2.8cm}=-\frac{N_c\left(N_c^2-1\right)}{4\pi^6N_fx^8}\left(\frac{1}{3}-\xi\right)\log\Lambda^2
\end{align*}
Because of the way the colour traces are arranged, again we see that four of the remaining diagrams will automatically cancel. We are left with
\begin{align*}
&\begin{tikzpicture}[scale=0.4]
\draw[scalar] (0,0) (0:2) arc (0:180:2 and 1);
\draw[scalar] (0,0) (-180:2) arc (-180:0:2 and 1);
\draw[scalar] (0,0) (0:2) arc (0:180:2);
\draw[scalar] (0,0) (-180:2) arc (-180:0:2);
\draw[thick,fill] (-2,0) circle (4pt);
\draw[thick,fill] (2,0) circle (4pt);
\draw[gluon,red] (0,2) to (0,-2); 
\node at (-3,0) {$2\,\times\,$};
\node at (16.5,0) {$=2\mathop{\mathlarger{\int}} dz\,dw\,D_{\mu\nu}^{AB}(z,w)\text{Tr}[G_a^{x_1}(x,z)\gamma^\mu (T_A)_{x_1}^{x_2}G_{x_2}^d(z,0)G_c^b(0,x)]$};
\end{tikzpicture}\\
&\hspace{3.1cm}\times \text{Tr}[G_b^c(x,0)G_{d}^{y_1}(0,w)\gamma^\nu(T_B)_{y_1}^{y_2}G_{y_2}^{a}(w,x)]\\
&\hspace{2.7cm}\simeq-\frac{4N_c\left(N_c^2-1\right)}{(4\pi)^4x^8}\int dz\,dw\, D_{\mu\nu}(z,w)\text{Tr}[\gamma^\mu G(z,0)]\,\text{Tr}[G(0,w)\gamma^\nu]\\
&\hspace{2.7cm}=\frac{N_c\left(N_c^2-1\right)}{4\pi^6N_fx^8}(1-\xi)\log\Lambda^2
\end{align*}
In total,
\begin{align}
\boxed{\langle O_1^s(x)\bar{O}_1^s(0) \rangle=\langle O_2^s(x)\bar{O}_2^s(0) \rangle=\frac{N_c\left(N_c^2-1\right)}{6\pi^6N_fx^8}\log\Lambda^2}
\end{align}

\subsubsection*{$\langle O_1^s(x)\bar{O}_2^s(0) \rangle$}

The calculation is completely analogous to the previous ones, so from this point on we just report the results of the diagrams.
\begin{align*}
&\begin{tikzpicture}[scale=0.4]
\draw[scalar] (0,0) (0:2) arc (0:180:2 and 1);
\draw[scalar] (0,0) (-180:2) arc (-180:0:2 and 1);
\draw[scalar] (0,0) (0:2) arc (0:180:2);
\draw[scalar] (0,0) (-180:2) arc (-180:0:2);
\draw[thick,fill] (-2,0) circle (4pt);
\draw[thick,fill] (2,0) circle (4pt);
\draw[gluon,red] (-1.37,1.4) to (1.37,1.4); 
\node at (-3,0) {$4\,\times\,$};
\node at (8.3,0) {$=\frac{N_c^2-1}{8\pi^6N_fx^8}\left(\frac{1}{3}-\xi\right)\log\Lambda^2  $};
\end{tikzpicture}\\
&\begin{tikzpicture}[scale=0.4]
\draw[scalar] (0,0) (0:2) arc (0:180:2 and 1);
\draw[scalar] (0,0) (-180:2) arc (-180:0:2 and 1);
\draw[scalar] (0,0) (0:2) arc (0:180:2);
\draw[scalar] (0,0) (-180:2) arc (-180:0:2);
\draw[thick,fill] (-2,0) circle (4pt);
\draw[thick,fill] (2,0) circle (4pt);
\draw[gluon,red] (0,2) to (0,1); 
\node at (-3,0) {$2\,\times\,$};
\node at (8.5,0) {$=-\frac{N_c^2-1}{8\pi^6N_fx^8}(1+\xi)\log\Lambda^2$};
\end{tikzpicture}\\
&\begin{tikzpicture}[scale=0.4]
\draw[scalar] (0,0) (0:2) arc (0:180:2 and 1);
\draw[scalar] (0,0) (-180:2) arc (-180:0:2 and 1);
\draw[scalar] (0,0) (0:2) arc (0:180:2);
\draw[scalar] (0,0) (-180:2) arc (-180:0:2);
\draw[thick,fill] (-2,0) circle (4pt);
\draw[thick,fill] (2,0) circle (4pt);
\draw[gluon,red] (0,2) to (0,-1); 
\node at (-3,0) {$4\,\times\,$};
\node at (8.5,0) {$=-\frac{N_c^2-1}{4\pi^6N_fx^8}(3-\xi)\log\Lambda^2$};
\end{tikzpicture}
\end{align*}
In total,
\begin{align}
\boxed{\langle O_1^s(x)\bar{O}_2^s(0) \rangle = -\frac{5\left(N_c^2-1\right)}{6\pi^6N_fx^8}\log\Lambda^2}
\end{align}

\subsubsection*{Mixed operators}

\subsubsection*{$\langle O_1^d(x)\bar{O}_1^s(0) \rangle=\langle O_2^d(x)\bar{O}_2^s(0) \rangle$}

\begin{align*}
&\begin{tikzpicture}[scale=0.4]
\draw[scalar] (0,0) (0:2) arc (0:180:2 and 1);
\draw[scalar] (0,0) (-180:2) arc (-180:0:2 and 1);
\draw[scalar] (0,0) (0:2) arc (0:180:2);
\draw[scalar] (0,0) (-180:2) arc (-180:0:2);
\draw[thick,fill] (-2,0) circle (4pt);
\draw[thick,fill] (2,0) circle (4pt);
\draw[gluon,red] (-1.37,1.4) to (1.37,1.4); 
\node at (-3,0) {$4\,\times\,$};
\node at (8.65,0) {$=-\frac{N_c^2-1}{4\pi^6N_fx^8}\left(\frac{1}{3}-\xi\right)\log\Lambda^2  $};
\end{tikzpicture}\\
&\begin{tikzpicture}[scale=0.4]
\draw[scalar] (0,0) (0:2) arc (0:180:2 and 1);
\draw[scalar] (0,0) (-180:2) arc (-180:0:2 and 1);
\draw[scalar] (0,0) (0:2) arc (0:180:2);
\draw[scalar] (0,0) (-180:2) arc (-180:0:2);
\draw[thick,fill] (-2,0) circle (4pt);
\draw[thick,fill] (2,0) circle (4pt);
\draw[gluon,red] (0,2) to (0,-1); 
\node at (-3,0) {$2\,\times\,$};
\node at (8,0) {$=\frac{N_c^2-1}{4\pi^6N_fx^8}(3-\xi)\log\Lambda^2$};
\end{tikzpicture}
\end{align*}
The four remaining diagrams cancel pairwise (the only difference in their expressions is $G(0,w)\to G(w,0)$). Thus, in total,
\begin{align}
\boxed{\langle O_1^d(x)\bar{O}_1^s(0) \rangle=\langle O_2^d(x)\bar{O}_2^s(0) \rangle = \frac{2\left(N_c^2-1\right)}{3\pi^6N_fx^8}\log\Lambda^2}
\end{align}

\subsubsection*{$\langle O_1^d(x)\bar{O}_2^s(0) \rangle=\langle O_2^d(x)\bar{O}_1^s(0) \rangle$}

\begin{align*}
&\begin{tikzpicture}[scale=0.4]
\draw[scalar] (0,0) (0:2) arc (0:180:2 and 1);
\draw[scalar] (0,0) (-180:2) arc (-180:0:2 and 1);
\draw[scalar] (0,0) (0:2) arc (0:180:2);
\draw[scalar] (0,0) (-180:2) arc (-180:0:2);
\draw[thick,fill] (-2,0) circle (4pt);
\draw[thick,fill] (2,0) circle (4pt);
\draw[gluon,red] (-1.37,1.4) to (1.37,1.4); 
\node at (-3,0) {$4\,\times\,$};
\node at (8.3,0) {$=\frac{N_c(N_c^2-1)}{8\pi^6N_fx^8}\left(\frac{1}{3}-\xi\right)\log\Lambda^2  $};
\end{tikzpicture}\\
&\begin{tikzpicture}[scale=0.4]
\draw[scalar] (0,0) (0:2) arc (0:180:2 and 1);
\draw[scalar] (0,0) (-180:2) arc (-180:0:2 and 1);
\draw[scalar] (0,0) (0:2) arc (0:180:2);
\draw[scalar] (0,0) (-180:2) arc (-180:0:2);
\draw[thick,fill] (-2,0) circle (4pt);
\draw[thick,fill] (2,0) circle (4pt);
\draw[gluon,red] (0,2) to (0,-1); 
\node at (-3,0) {$2\,\times\,$};
\node at (8.5,0) {$=-\frac{N_c(N_c^2-1)}{8\pi^6N_fx^8}(3-\xi)\log\Lambda^2$};
\end{tikzpicture}
\end{align*}
In this occasion, the remaining four diagrams vanish, again due to Tr$[T_A]=0$. Therefore,
\begin{align}
\boxed{\langle O_1^d(x)\bar{O}_2^s(0) \rangle=\langle O_2^d(x)\bar{O}_1^s(0) \rangle = -\frac{N_c\left(N_c^2-1\right)}{3\pi^6N_fx^8}\log\Lambda^2}
\end{align}

\end{appendix}

\bibliographystyle{JHEP}
\bibliography{ref}

\providecommand{\href}[2]{#2}\begingroup\raggedright\begin{thebibliography}{10}

\bibitem{PhysRevB.65.165113}
X.-G. Wen, \emph{Quantum orders and symmetric spin liquids},
  \href{https://doi.org/10.1103/PhysRevB.65.165113}{\emph{Phys. Rev. B}
  {\bfseries 65} (2002) 165113}.

\bibitem{Komargodski:2017keh}
Z.~Komargodski and N.~Seiberg, \emph{{A symmetry breaking scenario for
  QCD$_{3}$}}, \href{https://doi.org/10.1007/JHEP01(2018)109}{\emph{JHEP}
  {\bfseries 01} (2018) 109}
  [\href{https://arxiv.org/abs/1706.08755}{{\ttfamily 1706.08755}}].

\bibitem{Gaiotto:2017tne}
D.~Gaiotto, Z.~Komargodski and N.~Seiberg, \emph{{Time-reversal breaking in
  QCD$_{4}$, walls, and dualities in 2 + 1 dimensions}},
  \href{https://doi.org/10.1007/JHEP01(2018)110}{\emph{JHEP} {\bfseries 01}
  (2018) 110} [\href{https://arxiv.org/abs/1708.06806}{{\ttfamily
  1708.06806}}].

\bibitem{Armoni:2019lgb}
A.~Armoni, T.~T. Dumitrescu, G.~Festuccia and Z.~Komargodski, \emph{{Metastable
  vacua in large-N QCD$_{3}$}},
  \href{https://doi.org/10.1007/JHEP01(2020)004}{\emph{JHEP} {\bfseries 01}
  (2020) 004} [\href{https://arxiv.org/abs/1905.01797}{{\ttfamily
  1905.01797}}].

\bibitem{Argurio:2019tvw}
R.~Argurio, M.~Bertolini, F.~Mignosa and P.~Niro, \emph{{Charting the phase
  diagram of QCD$_{3}$}},
  \href{https://doi.org/10.1007/JHEP08(2019)153}{\emph{JHEP} {\bfseries 08}
  (2019) 153} [\href{https://arxiv.org/abs/1905.01460}{{\ttfamily
  1905.01460}}].

\bibitem{Argurio:2020her}
R.~Argurio, A.~Armoni, M.~Bertolini, F.~Mignosa and P.~Niro, \emph{{Vacuum
  structure of large $N$ $QCD_{3}$ from holography}},
  \href{https://doi.org/10.1007/JHEP07(2020)134}{\emph{JHEP} {\bfseries 07}
  (2020) 134} [\href{https://arxiv.org/abs/2006.01755}{{\ttfamily
  2006.01755}}].

\bibitem{Benvenuti:2019ujm}
S.~Benvenuti and H.~Khachatryan, \emph{{Easy-plane QED$_{3}$\textquoteright{}s
  in the large N$_{f}$ limit}},
  \href{https://doi.org/10.1007/JHEP05(2019)214}{\emph{JHEP} {\bfseries 05}
  (2019) 214} [\href{https://arxiv.org/abs/1902.05767}{{\ttfamily
  1902.05767}}].

\bibitem{Gracey:2018fwq}
J.~A. Gracey, \emph{{Fermion bilinear operator critical exponents at $O(1/N^2)$
  in the QED-Gross-Neveu universality class}},
  \href{https://doi.org/10.1103/PhysRevD.98.085012}{\emph{Phys. Rev. D}
  {\bfseries 98} (2018) 085012}
  [\href{https://arxiv.org/abs/1808.07697}{{\ttfamily 1808.07697}}].

\bibitem{Pisarski:1984dj}
R.~D. Pisarski, \emph{{Chiral Symmetry Breaking in Three-Dimensional
  Electrodynamics}},
  \href{https://doi.org/10.1103/PhysRevD.29.2423}{\emph{Phys. Rev. D}
  {\bfseries 29} (1984) 2423}.

\bibitem{Appelquist:1986fd}
T.~W. Appelquist, M.~J. Bowick, D.~Karabali and L.~C.~R. Wijewardhana,
  \emph{{Spontaneous Chiral Symmetry Breaking in Three-Dimensional QED}},
  \href{https://doi.org/10.1103/PhysRevD.33.3704}{\emph{Phys. Rev. D}
  {\bfseries 33} (1986) 3704}.

\bibitem{Appelquist:1988sr}
T.~Appelquist, D.~Nash and L.~C.~R. Wijewardhana, \emph{{Critical Behavior in
  (2+1)-Dimensional QED}},
  \href{https://doi.org/10.1103/PhysRevLett.60.2575}{\emph{Phys. Rev. Lett.}
  {\bfseries 60} (1988) 2575}.

\bibitem{Kaveh:2004qa}
K.~Kaveh and I.~F. Herbut, \emph{{Chiral symmetry breaking in QED(3) in
  presence of irrelevant interactions: A Renormalization group study}},
  \href{https://doi.org/10.1103/PhysRevB.71.184519}{\emph{Phys. Rev. B}
  {\bfseries 71} (2005) 184519}
  [\href{https://arxiv.org/abs/cond-mat/0411594}{{\ttfamily
  cond-mat/0411594}}].

\bibitem{Braun:2014wja}
J.~Braun, H.~Gies, L.~Janssen and D.~Roscher, \emph{{Phase structure of
  many-flavor QED$_3$}},
  \href{https://doi.org/10.1103/PhysRevD.90.036002}{\emph{Phys. Rev. D}
  {\bfseries 90} (2014) 036002}
  [\href{https://arxiv.org/abs/1404.1362}{{\ttfamily 1404.1362}}].

\bibitem{DiPietro:2015taa}
L.~Di~Pietro, Z.~Komargodski, I.~Shamir and E.~Stamou, \emph{{Quantum
  Electrodynamics in d=3 from the \ensuremath{\varepsilon} Expansion}},
  \href{https://doi.org/10.1103/PhysRevLett.116.131601}{\emph{Phys. Rev. Lett.}
  {\bfseries 116} (2016) 131601}
  [\href{https://arxiv.org/abs/1508.06278}{{\ttfamily 1508.06278}}].

\bibitem{Herbut:2016ide}
I.~F. Herbut, \emph{{Chiral symmetry breaking in three-dimensional quantum
  electrodynamics as fixed point annihilation}},
  \href{https://doi.org/10.1103/PhysRevD.94.025036}{\emph{Phys. Rev. D}
  {\bfseries 94} (2016) 025036}
  [\href{https://arxiv.org/abs/1605.09482}{{\ttfamily 1605.09482}}].

\bibitem{Benvenuti:2018cwd}
S.~Benvenuti and H.~Khachatryan, \emph{{QED's in $2{+}1$ dimensions: complex
  fixed points and dualities}},
  \href{https://arxiv.org/abs/1812.01544}{{\ttfamily 1812.01544}}.

\bibitem{Kaplan:2010zz}
D.~B. Kaplan, J.~W. Lee, D.~T. Son and M.~A. Stephanov, \emph{{Conformality
  lost}}, \href{https://doi.org/10.1142/S0217751X1004872X}{\emph{Int. J. Mod.
  Phys. A} {\bfseries 25} (2010) 422}.

\bibitem{Gukov:2016tnp}
S.~Gukov, \emph{{RG Flows and Bifurcations}},
  \href{https://doi.org/10.1016/j.nuclphysb.2017.03.025}{\emph{Nucl. Phys. B}
  {\bfseries 919} (2017) 583}
  [\href{https://arxiv.org/abs/1608.06638}{{\ttfamily 1608.06638}}].

\bibitem{Kuipers:2018lux}
F.~Kuipers, U.~G\"ursoy and Y.~Kuznetsov, \emph{{Bifurcations in the RG-flow of
  QCD}}, \href{https://doi.org/10.1007/JHEP07(2019)075}{\emph{JHEP} {\bfseries
  07} (2019) 075} [\href{https://arxiv.org/abs/1812.05179}{{\ttfamily
  1812.05179}}].

\bibitem{Appelquist:1989tc}
T.~Appelquist and D.~Nash, \emph{{Critical Behavior in (2+1)-dimensional
  {QCD}}}, \href{https://doi.org/10.1103/PhysRevLett.64.721}{\emph{Phys. Rev.
  Lett.} {\bfseries 64} (1990) 721}.

\bibitem{Gies:2005as}
H.~Gies and J.~Jaeckel, \emph{{Chiral phase structure of QCD with many
  flavors}}, \href{https://doi.org/10.1140/epjc/s2006-02475-0}{\emph{Eur. Phys.
  J. C} {\bfseries 46} (2006) 433}
  [\href{https://arxiv.org/abs/hep-ph/0507171}{{\ttfamily hep-ph/0507171}}].

\bibitem{PhysRevB.78.054432}
C.~Xu, \emph{Renormalization group studies on four-fermion interaction
  instabilities on algebraic spin liquids},
  \href{https://doi.org/10.1103/PhysRevB.78.054432}{\emph{Phys. Rev. B}
  {\bfseries 78} (2008) 054432}.

\bibitem{Goldman:2016wwk}
H.~Goldman and M.~Mulligan, \emph{{Stability of $SU(N_c)$ QCD$_3$ from the
  $\epsilon$-Expansion}},
  \href{https://doi.org/10.1103/PhysRevD.94.065031}{\emph{Phys. Rev. D}
  {\bfseries 94} (2016) 065031}
  [\href{https://arxiv.org/abs/1606.07067}{{\ttfamily 1606.07067}}].

\bibitem{Gomis:2017ixy}
J.~Gomis, Z.~Komargodski and N.~Seiberg, \emph{{Phases Of Adjoint QCD$_3$ And
  Dualities}},
  \href{https://doi.org/10.21468/SciPostPhys.5.1.007}{\emph{SciPost Phys.}
  {\bfseries 5} (2018) 007} [\href{https://arxiv.org/abs/1710.03258}{{\ttfamily
  1710.03258}}].

\bibitem{Choi:2018tuh}
C.~Choi, D.~Delmastro, J.~Gomis and Z.~Komargodski, \emph{{Dynamics of
  QCD$_{3}$ with Rank-Two Quarks And Duality}},
  \href{https://doi.org/10.1007/JHEP03(2020)078}{\emph{JHEP} {\bfseries 03}
  (2020) 078} [\href{https://arxiv.org/abs/1810.07720}{{\ttfamily
  1810.07720}}].

\bibitem{Choi:2019eyl}
C.~Choi, \emph{{Phases of Two Adjoints QCD$_{3}$ And a Duality Chain}},
  \href{https://doi.org/10.1007/JHEP04(2020)006}{\emph{JHEP} {\bfseries 04}
  (2020) 006} [\href{https://arxiv.org/abs/1910.05402}{{\ttfamily
  1910.05402}}].

\bibitem{Tanizaki:2018wtg}
Y.~Tanizaki, \emph{{Anomaly constraint on massless QCD and the role of
  Skyrmions in chiral symmetry breaking}},
  \href{https://doi.org/10.1007/JHEP08(2018)171}{\emph{JHEP} {\bfseries 08}
  (2018) 171} [\href{https://arxiv.org/abs/1807.07666}{{\ttfamily
  1807.07666}}].

\bibitem{Armoni:2017jkl}
A.~Armoni and V.~Niarchos, \emph{{Phases of QCD$_3$ from Non-SUSY Seiberg
  Duality and Brane Dynamics}},
  \href{https://doi.org/10.1103/PhysRevD.97.106001}{\emph{Phys. Rev. D}
  {\bfseries 97} (2018) 106001}
  [\href{https://arxiv.org/abs/1711.04832}{{\ttfamily 1711.04832}}].

\bibitem{Lohitsiri:2022jyz}
N.~Lohitsiri and T.~Sulejmanpasic, \emph{{Comments on QCD$_{3}$ and anomalies
  with fundamental and adjoint matter}},
  \href{https://doi.org/10.1007/JHEP10(2022)081}{\emph{JHEP} {\bfseries 10}
  (2022) 081} [\href{https://arxiv.org/abs/2205.07825}{{\ttfamily
  2205.07825}}].

\bibitem{Bashmakov:2021rci}
V.~Bashmakov and N.~Gorini, \emph{{Phases of $ \mathcal{N} $ = 1 quivers in 2 +
  1 dimensions}}, \href{https://doi.org/10.1007/JHEP07(2022)110}{\emph{JHEP}
  {\bfseries 07} (2022) 110}
  [\href{https://arxiv.org/abs/2109.11862}{{\ttfamily 2109.11862}}].

\bibitem{Delmastro:2021otj}
D.~Delmastro, J.~Gomis and M.~Yu, \emph{{Infrared phases of 2d QCD}},
  \href{https://arxiv.org/abs/2108.02202}{{\ttfamily 2108.02202}}.

\bibitem{Kan:2019rsz}
N.~Kan, R.~Kitano, S.~Yankielowicz and R.~Yokokura, \emph{{From 3d dualities to
  hadron physics}},
  \href{https://doi.org/10.1103/PhysRevD.102.125034}{\emph{Phys. Rev. D}
  {\bfseries 102} (2020) 125034}
  [\href{https://arxiv.org/abs/1909.04082}{{\ttfamily 1909.04082}}].

\bibitem{Aitken:2019mtq}
K.~Aitken, A.~Baumgartner, C.~Choi and A.~Karch, \emph{{Generalization of
  QCD$_{3}$ symmetry-breaking and flavored quiver dualities}},
  \href{https://doi.org/10.1007/JHEP02(2020)060}{\emph{JHEP} {\bfseries 02}
  (2020) 060} [\href{https://arxiv.org/abs/1906.08785}{{\ttfamily
  1906.08785}}].

\bibitem{Akhond:2019ued}
M.~Akhond, A.~Armoni and S.~Speziali, \emph{{Phases of $\mathbf{U(N_c)}$
  QCD$_3$ from Type 0 Strings and Seiberg Duality}},
  \href{https://doi.org/10.1007/JHEP09(2019)111}{\emph{JHEP} {\bfseries 09}
  (2019) 111} [\href{https://arxiv.org/abs/1908.04324}{{\ttfamily
  1908.04324}}].

\bibitem{Amariti:2018wht}
A.~Amariti and L.~Cassia, \emph{{USp(2N$_{c}$) SQCD$_{3}$ with antisymmetric:
  dualities and symmetry enhancements}},
  \href{https://doi.org/10.1007/JHEP02(2019)013}{\emph{JHEP} {\bfseries 02}
  (2019) 013} [\href{https://arxiv.org/abs/1809.03796}{{\ttfamily
  1809.03796}}].

\bibitem{Dey:2018ykx}
A.~Dey, I.~Halder, S.~Jain, L.~Janagal, S.~Minwalla and N.~Prabhakar,
  \emph{{Duality and an exact Landau-Ginzburg potential for quasi-bosonic
  Chern-Simons-Matter theories}},
  \href{https://doi.org/10.1007/JHEP11(2018)020}{\emph{JHEP} {\bfseries 11}
  (2018) 020} [\href{https://arxiv.org/abs/1808.04415}{{\ttfamily
  1808.04415}}].

\bibitem{Aharony:2018pjn}
O.~Aharony, S.~Jain and S.~Minwalla, \emph{{Flows, Fixed Points and Duality in
  Chern-Simons-matter theories}},
  \href{https://doi.org/10.1007/JHEP12(2018)058}{\emph{JHEP} {\bfseries 12}
  (2018) 058} [\href{https://arxiv.org/abs/1808.03317}{{\ttfamily
  1808.03317}}].

\bibitem{Choi:2018ohn}
C.~Choi, M.~Ro\v{c}ek and A.~Sharon, \emph{{Dualities and Phases of $3D N=1$
  SQCD}}, \href{https://doi.org/10.1007/JHEP10(2018)105}{\emph{JHEP} {\bfseries
  10} (2018) 105} [\href{https://arxiv.org/abs/1808.02184}{{\ttfamily
  1808.02184}}].

\bibitem{Gaiotto:2018yjh}
D.~Gaiotto, Z.~Komargodski and J.~Wu, \emph{{Curious Aspects of
  Three-Dimensional ${\cal N}=1$ SCFTs}},
  \href{https://doi.org/10.1007/JHEP08(2018)004}{\emph{JHEP} {\bfseries 08}
  (2018) 004} [\href{https://arxiv.org/abs/1804.02018}{{\ttfamily
  1804.02018}}].

\bibitem{Benini:2018umh}
F.~Benini and S.~Benvenuti, \emph{{$ \mathcal{N} $ = 1 dualities in 2+1
  dimensions}}, \href{https://doi.org/10.1007/JHEP11(2018)197}{\emph{JHEP}
  {\bfseries 11} (2018) 197}
  [\href{https://arxiv.org/abs/1803.01784}{{\ttfamily 1803.01784}}].

\bibitem{Bashmakov:2018wts}
V.~Bashmakov, J.~Gomis, Z.~Komargodski and A.~Sharon, \emph{{Phases of $
  \mathcal{N}=1 $ theories in 2 + 1 dimensions}},
  \href{https://doi.org/10.1007/JHEP07(2018)123}{\emph{JHEP} {\bfseries 07}
  (2018) 123} [\href{https://arxiv.org/abs/1802.10130}{{\ttfamily
  1802.10130}}].

\bibitem{Cordova:2017kue}
C.~C\'ordova, P.-S. Hsin and N.~Seiberg, \emph{{Time-Reversal Symmetry,
  Anomalies, and Dualities in (2+1)$d$}},
  \href{https://doi.org/10.21468/SciPostPhys.5.1.006}{\emph{SciPost Phys.}
  {\bfseries 5} (2018) 006} [\href{https://arxiv.org/abs/1712.08639}{{\ttfamily
  1712.08639}}].

\bibitem{Chester:2016ref}
S.~M. Chester and S.~S. Pufu, \emph{{Anomalous dimensions of scalar operators
  in QED$_{3}$}}, \href{https://doi.org/10.1007/JHEP08(2016)069}{\emph{JHEP}
  {\bfseries 08} (2016) 069}
  [\href{https://arxiv.org/abs/1603.05582}{{\ttfamily 1603.05582}}].

\bibitem{Lee:2018udi}
J.~Y. Lee, C.~Wang, M.~P. Zaletel, A.~Vishwanath and Y.-C. He, \emph{{Emergent
  Multi-flavor QED3 at the Plateau Transition between Fractional Chern
  Insulators: Applications to graphene heterostructures}},
  \href{https://doi.org/10.1103/PhysRevX.8.031015}{\emph{Phys. Rev. X}
  {\bfseries 8} (2018) 031015}
  [\href{https://arxiv.org/abs/1802.09538}{{\ttfamily 1802.09538}}].

\bibitem{Rantner:2002zz}
W.~Rantner and X.-G. Wen, \emph{{Spin correlations in the algebraic spin
  liquid: Implications for high-Tc superconductors}},
  \href{https://doi.org/10.1103/PhysRevB.66.144501}{\emph{Phys. Rev. B}
  {\bfseries 66} (2002) 144501}
  [\href{https://arxiv.org/abs/cond-mat/0201521}{{\ttfamily
  cond-mat/0201521}}].

\end{thebibliography}\endgroup

\end{document}